\documentclass[12pt]{article}
\usepackage{epsfig}
\usepackage{latexsym}
\usepackage{cite}

\usepackage{pslatex}
\usepackage[latin1]{inputenc}
\usepackage[T1]{fontenc}

\usepackage{color,colordvi}
\def\colour4colour#1{\Blue{#1}}

\newcommand{\as}{\alpha_{\rm s}}
\newcommand{\ar}{a_{\rm s}}

\def\GeV{\rm{GeV}}
\def\MSbar{\overline{\mathrm{MS}}}

\def\z#1{{\zeta_{#1}}}
\def\ca{{C^{}_A}}
\def\cf{{C^{}_F}}
\def\nf{{n^{}_{\! f}}}

\def\S(#1){{{S}_{#1}}}
\def\Ss(#1,#2){{{S}_{#1,#2}}}
\def\Sss(#1,#2,#3){{{S}_{#1,#2,#3}}}
\def\Ssss(#1,#2,#3,#4){{{S}_{#1,#2,#3,#4}}}
\def\Sssss(#1,#2,#3,#4,#5){{{S}_{#1,#2,#3,#4,#5}}}
\def\Npm{{{\bf N_{\pm}}}}
\def\Npmi{{{\bf N_{\pm i}}}}
\def\Nminus{{{\bf N_{-}}}}
\def\Nplus{{{\bf N_{+}}}}
\def\Nminustwo{{{\bf N_{-2}}}}
\def\Nplustwo{{{\bf N_{+2}}}}
\def\Nminusthree{{{\bf N_{-3}}}}
\def\Nplusthree{{{\bf N_{+3}}}}

\def\pqq(#1){p_{\rm{qq}}(#1)}
\def\pgg(#1){p_{\rm{gg}}(#1)}
\def\H(#1){{\rm{H}}_{#1}}
\def\Hh(#1,#2){{\rm{H}}_{#1,#2}}
\def\Hhh(#1,#2,#3){{\rm{H}}_{#1,#2,#3}}
\def\Hhhh(#1,#2,#3,#4){{\rm{H}}_{#1,#2,#3,#4}}

\begin{document}
\setlength{\parskip}{0.2cm} \setlength{\baselineskip}{0.55cm}

\begin{titlepage}
\noindent
DESY 06-043 \hfill {\tt hep-ph/0604160}\\
April 2006 \\
\vspace{1.8cm}
\begin{center}
\LARGE {\bf QCD Corrections to Semi-Inclusive Hadron \\
            Production in Electron-Positron Annihilation \\
            at Two Loops} \\
\vspace{2.2cm}
\large
A. Mitov and S. Moch \\
\vspace{1.4cm}
\normalsize
{\it Deutsches Elektronensynchrotron DESY \\
\vspace{0.1cm}
Platanenallee 6, D--15738 Zeuthen, Germany}\\
\vfill
\large {\bf Abstract}
\vspace{-0.2cm}
\end{center}
We report on the analytic calculation of the second-order QCD
corrections to semi-inclusive hadron production in
electron-positron annihilation. The calculation of the
single-particle inclusive cross-section in time-like kinematics
is performed in Mellin $N$-space and uses an algorithmic evaluation
of inclusive phase-space integrals, based on the unitarity cutting rules
and integration-by-parts.
We obtain splitting functions and coefficient functions up to second order
in the strong coupling $\alpha_{\rm s}$.
Our results are in agreement with earlier calculations
in the literature by Rijken and van Neerven.
\\
\vspace{3.0cm}
\end{titlepage}

%
%
\section{Introduction}
\label{sec:intro}

The direct observation of hadron production in $e^+ e^-$-annihilation
offers unique ways to test predictions of QCD and has
been measured in the past especially by the LEP experiments, see
Ref.~\cite{Biebel:2001ka,Kluth:2006bw}. These data allow in
particular for studies of single hadron production, fragmentation
functions and their scaling violations.
They also offer the possibility for determinations of
the strong coupling constant $\as$ at the scale $M_Z$.

In parallel to the deep-inelastic scattering (DIS) of leptons on
hadrons, the fragmentation of the partons produced in the hard
scattering process $e^+e^-\rightarrow q,{\bar q}, g$ into hadrons
depends on the hadron's scaled momentum $x$. Perturbative QCD
predicts the scale dependence and evolution of the scattering
process, making higher order corrections mandatory for precision
analyses. In the past, the next-to-leading order (NLO) results
have been obtained in
Refs.~\cite{Curci:1980uw,Furmanski:1980cm,Floratos:1981hs,Nason:1993xx},
while the coefficient functions at two loops, necessary for a
next-to-next-to-leading order (NNLO) analysis were
calculated~\cite{Rijken:1996vr,Rijken:1996np,Rijken:1996ns}. In
order to complete this program, the three-loop QCD predictions for
the splitting function in time-like kinematics are required. The
splitting functions governing the NNLO evolution of flavor
non-singlet fragmentation distributions at third-order have
recently been reported~\cite{Mitov:2006ic}, while the
corresponding singlet quantities are still missing.

Our motivation for the present paper is at least two-fold. First
of all, given that $e^+ e^-$-annihilation is of particular
interest for the physics analyses of LEP data or a future ILC,
we would like to provide an independent cross-check on the computation
of Rijken and van Neerven~\cite{Rijken:1996vr,Rijken:1996np,Rijken:1996ns}.
Secondly, we perform a first application of a new innovative
method~\cite{Mitov:2005ps} to calculate higher order QCD
corrections to single-particle inclusive observables directly in
Mellin $N$-space. It is worth emphasizing that the present method
extends well beyond the framework of the operator product
expansion (OPE) used in
DIS~\cite{Kazakov:1987jk,Moch:1999eb,Moch:2004pa,Vogt:2004mw,Vermaseren:2005qc}.
This latter point is, in a more general perspective, rather
important as it will allow for the efficient calculation of QCD
corrections to many single scale observables.
This can be done for Mellin moments either at fixed $N$ analogous to
DIS calculations of Refs.~\cite{Larin:1994vu,Larin:1997wd,Retey:2000nq,Blumlein:2004xt}
or by means of well established summation techniques~\cite{Moch:2005uq}
when full control over the analytic $N$ dependence is kept.

Finally, the present paper provides the means to thoroughly study
the relation between the time-like QCD corrections to inclusive
hadron production in $e^+e^-$ annihilation and their space-like
counterparts, the DIS structure functions. At the leading order
(LO) the Gribov-Lipatov relation~\cite{Gribov:1972ri} suggests
simple relations between the splitting functions in both
kinematics, which do not hold beyond
LO~\cite{Curci:1980uw,Stratmann:1996hn}, see also
Refs.~\cite{Broadhurst:1993ru,Brodsky:1995tb,Blumlein:2000wh}.
Yet, the space- and time-like cases are related by an analytic
continuation in $x$ which has recently been used to obtain the
NNLO flavor non-singlet time-like splitting
functions~\cite{Mitov:2006ic}. In the present paper, we have also
calculated in $d=4-2\epsilon$ dimensions so-far unknown terms at
higher orders in $\epsilon$ for the corresponding coefficient
functions in $e^+e^-$-annihilation. These terms were subsequently
used to check the analytical continuation between processes with
space- and time-like kinematics~\cite{Mitov:2006ic}.

The outline of the article is as follows. In
Section~\ref{sec:setup} we set up the notation and the formalism
for calculating QCD corrections to fragmentation functions. We
also discuss the definition of the (time-like) splitting functions
and respective coefficient functions through order $\as^2$.
Section~\ref{sec:method} briefly explains the method to calculate
Feynman diagrams in Mellin $N$-space for processes with time-like
kinematics, with all details on the necessary master integrals and
reductions given in Appendix A. In Section~\ref{sec:results} we
present our results for the fragmentation functions $F_T, F_L$ and
$F_A$ through order $\as^2$ and with details on the mass
factorization to extract the splitting and coefficient functions.
The lengthy full expressions are deferred to Appendix B
($N$-space) and Appendix C ($x$-space). We summarize in
Section~\ref{sec:summary}.

%
%
\section{The setup}
\label{sec:setup}

The subject of our calculation is the single-particle inclusive
$e^+ e^-$-annihilation, i.e. the process
\begin{equation}
\label{eq:epem}
e^+(k_1) + e^-(k_2) \:\:\rightarrow\:\: V(q) \:\:\rightarrow\:\: H(p_H) \:+\:  X \, ,
\end{equation}
where $V$ is a vector boson, i.e. $V=\gamma,Z$, and $(k_1+k_2)^2 =
q^2 = Q^2>0$ is its (time-like) four-momentum squared. The
observed hadron with momentum $p_H$ is denoted by $H$ and $X$
stands for any hadronic final states allowed by quantum
number conservation.

Our goal is the derivation of the normalized double differential
cross-section for the reaction in Eq.~(\ref{eq:epem}),
\begin{equation}
\label{eq:doublediffcrs}
 {1 \over \sigma_{\rm tot}}\, {d^2 \sigma^H \over dx \, d\cos\theta} =
   {3 \over 8}(1+\cos^2\theta)\, F_T(x) +
   {3 \over 4}\sin^2\theta\, F_L(x) +
   {3 \over 4}\cos\theta\, F_A(x)
   \, ,
\end{equation}
in perturbation theory, and including the quantum corrections
to the fragmentation functions $F_T, F_L$ and
$F_A$ through order ${\cal{O}}(\as^2)$ in the strong coupling.
In Eq.~(\ref{eq:doublediffcrs}) $\sigma^H$ denotes the cross-section
for producing the hadron $H$, $x$ is its scaled momentum
fraction:
\begin{equation}
  \label{eq:xdef}
x = {2 p_H \cdot q \over Q^2}\, , \quad\quad\quad  0 \le x \le 1\,
,
\end{equation}
and $\theta$ denotes the polar angle between the hadron and
electron beam directions.
The (total) fragmentation functions $F_T, F_L$ and $F_A$
in Eq.~(\ref{eq:doublediffcrs}) originate from the transverse or
longitudinal polarization states of the intermediate vector boson
($\gamma,Z$) or from parity violation of the electroweak interaction.
Upon integration over $\theta$ and $x$ the asymmetric
contribution $F_A$ cancels and one arrives at the total cross-section
integral
\begin{equation}
\label{eq:sigmatot}
  {1 \over 2}\int_0^1 dx\, x\,
  {1 \over \sigma_{\rm tot}}\, {d \sigma^H \over d x}
  \,=\,
  {\sigma_T \over \sigma_{\rm tot}} + {\sigma_L \over \sigma_{\rm tot}}
  \,=\, 1 \, ,
\end{equation}
which represents the energy-momentum sum-rule.

The evaluation of the fragmentation functions $F_T, F_L$ and $F_A$
in perturbative QCD is based on factorization. Up to power
corrections suppressed by the hard scale $Q$, one can write the
desired hadron level observables $F$ as a convolution of
{\it collinearly renormalized}, parton level cross-sections $\hat{\cal F}$
with non-perturbative fragmentation distributions $D$.
The explicit form of these relations is:
\begin{eqnarray}
  F_{I}(x) &=&
  \sigma_{tot}^{(0)}(Q^2)\left[
    \hat{\cal F}_{I,\, q} \otimes D^{{\rm s}}_{q\to H}
    + \hat{\cal F}_{I,\, g} \otimes D_{g\to H}\right]
  + \sum_{h=1}^{n_f}\sigma_h^{(0)}(Q^2)
  \hat{\cal F}_{I,\, {\rm ns}} \otimes D^{{\rm ns}, +}_{h\to H}\, ,
\nonumber\\
  F_{A}(x) &=&
  \sum_{h=1}^{n_f} A_h^{(0)}(Q^2)
  \hat{\cal F}_{A,\, {\rm {ns}}} \otimes D^{{\rm ns}, -}_{h\to H}\, ,
  \label{eq:dsigma}
\end{eqnarray}
where $I=T,L$ and the symbol $\otimes$ stands for the convolution integral,
\begin{equation}
  \label{eq:conv-def}
\left[ f\otimes g \right](z) = \int_0^1dx\int_0^1dy f(x) g(y) \delta(z-xy)\, .
\end{equation}
The factors $\sigma_{tot}^{(0)}$ and $A_{h}^{(0)}$
are respectively the LO total cross-section and asymmetry factor
for $e^+e^-\to hadrons$. Moreover,
\begin{equation}
  \label{eq:sigmatot-def}
  \sigma_{tot}^{(0)}(Q^2) = \sum_{h=1}^{n_f} \sigma_{h}^{(0)}(Q^2)\, .
\end{equation}
The explicit expressions
for $\sigma_{h}^{(0)}$ and $A_{h}^{(0)}$ can be found
in Ref.~\cite{Rijken:1996ns}.

The distributions $D$ for a non-singlet quark (of flavor $h$),
a singlet quark ($q$) or gluon $g$ fragmenting to the observed hadron $H$
are non-perturbative objects that are extracted from experimental data
at low scales $Q$ of order $1~\GeV$ and evolved to high scales
by means of the time-like Dokshitzer-Gribov-Lipatov-Altarelli-Parisi evolution
equations~\cite{Gribov:1972ri,Lipatov:1974qm,Dokshitzer:1977sg,Altarelli:1977zs}.

As discussed above, the time-like splitting functions are
presently known to NLO accuracy in the flavor singlet and to NNLO
in the non-singlet case~\cite{Mitov:2006ic}. Thus far, this has
permitted the determination of the fragmentation distributions with
next-to-leading logarithmic (NLL) accuracy. Examples of sets of LL
and NLL accuracy
are~\cite{Kretzer:2000yf,Bourhis:2000gs,Kniehl:2000fe,Kniehl:2000hk}.
In case the hadron $H$ contains a heavy flavor, e.g. $b$-quark,
the fragmentation distributions develop an additional perturbative
component, a so-called `perturbative fragmentation function' for the
heavy quark~\cite{Mele:1990cw}. That function allows the
resummation of large logarithms of the ratio of the quark mass and
the hard scale; it is currently known to
NNLO~\cite{Melnikov:2004bm,Mitov:2004du} which allows for the
extraction of heavy quark fragmentation distributions with NNLL
accuracy provided all three-loop time-like splitting functions are
available.

The relation between the fragmentation distributions introduced
above and the respective distributions for specific flavors are,
\begin{eqnarray}
\label{eq:D}
D^{{\rm ns}, +}_{h\to H} &=&
D_{h\to H} + D_{\bar{h}\to H} - D^{(S)}_{q\to H}\, ,
\nonumber\\
D^{{\rm ns}, -}_{h\to H} &=& D_{h\to H} - D_{\bar{h}\to H}\, ,
\nonumber\\
D^{{\rm s}}_{q\to H} &=&
{1\over n_f} \sum_{h=1}^{n_f}\left[ D_{h\to H} + D_{\bar{h}\to H} \right]\, ,
\end{eqnarray}
where $q$ denotes any generic `quark' flavor, while $h$ stands
for a specific quark flavor. The non-singlet distributions evolve with the
corresponding combinations of splitting functions~\cite{Mitov:2006ic}
(see also Eqs.~(30)--(34) in Ref.~\cite{Moch:1999eb}).

Next we turn our attention to the evaluation of the hard partonic
cross-sections $\hat{\cal F}$ appearing in Eq.~(\ref{eq:dsigma});
the explicit results for these functions are discussed in
Appendices B and C. The coefficient functions $\hat{\cal F}$ are
related to the partonic equivalents ${\cal F}$ of the
corresponding observables $F$. For the construction of the `bare'
functions ${\cal F}$ one replaces the observed hadron $H$ with an
on-shell massless parton that can be $q,\bar{q}$ or $g$. Clearly,
due to the incomplete inclusiveness of the
observable~(\ref{eq:epem}), the partonic cross-sections ${\cal F}$
will contain additional collinear singularities, i.e. these
functions are divergent even after the usual UV renormalization.
To that end one performs a so-called `collinear renormalization'
or `mass factorization'. One factorizes the collinear
singularities and effectively absorbs them into the fragmentation
distributions $D$. As an intermediate regularization we work in
$d=4-2\epsilon$
dimensions~\cite{'tHooft:1972fi,Bollini:1972ui,Ashmore:1972uj,Cicuta:1972jf}.
In $z$-space and in the $\MSbar$ scheme~\cite{Bardeen:1978yd},
these relation take the form (with $I=T,L$ and $J=T,L,A$):
\begin{eqnarray}
\label{eq:mass-factorization}
{\cal F}_{J,\, {\rm ns}}(\epsilon,z) &=& \hat{\cal F}_{J,\, {\rm ns}}(z) \otimes
\Gamma^{{\rm ns}}_{qq}(\epsilon,z), \nonumber\\
{\cal F}_{I,\, {\rm s}}(\epsilon,z) &=& \hat{\cal F}_{I,\, {\rm s}}(z) \otimes
\Gamma^{{\rm s}}_{qq}(\epsilon,z) + n_f \hat{\cal F}_{I,\, g}(z) \otimes
\Gamma_{qg}(\epsilon,z), \nonumber\\
{\cal F}_{I,\, g}(\epsilon,z) &=& 2 \hat{\cal F}_{I,\, q}(z) \otimes
\Gamma_{gq}(\epsilon,z) + \hat{\cal F}_{I,\, g}(z) \otimes
\Gamma_{gg}(\epsilon,z)\, ,
\end{eqnarray}
and in Section~\ref{sec:results} we explicitly present the
corresponding expression in $N$-space. The form of the collinear
counter-terms $\Gamma$ is universal, i.e. they contain only the
time-like splitting functions. In the $\MSbar$ scheme, and in
terms of the bare strong coupling, their explicit expressions can
be found in Eqs.~(4.25)--(4.30) of Ref.~\cite{Rijken:1996ns} and
we do not repeat them here\footnote{ We would only like to caution
the reader about the different notations for the definition of
$\epsilon$ and the normalization of the splitting functions used
in these papers. In addition, the conventions for these
counter-terms are the transposed of what is used in the more
current literature.}.

The partonic scaling variable $z$ appearing in
Eq.~(\ref{eq:mass-factorization}) is the normalized energy fraction of
the observed parton defined with respect to the total
four-momentum $q$ of the $e^+e^-$ system:
\begin{equation}
\label{eq:def-z}
z = {2 p \cdot q \over Q^2}\, , \quad\quad\quad  0 \le z \le 1\, .
\end{equation}
This definition of $z$ is consistent with the requirement that all
partons are massless (see below).

$\hat{\cal F}$ in Eq.~(\ref{eq:mass-factorization}) above are the
(finite) dimensionless partonic cross-sections with the collinear (or mass)
singularities factorized in the $\MSbar$ scheme. We have
suppressed the explicit dependence on the factorization scale
$\mu_F$; throughout this paper we set it equal to the hard scale
$\mu_F^2=Q^2$. Moreover, all functions appearing in
Eq.~(\ref{eq:mass-factorization}) are expressed in terms of the
renormalized strong coupling. The relation between the bare $\as$
and the renormalized $\as(\mu_R)$ couplings reads:
\begin{equation}
{\as\over 4\pi} S_\epsilon =
{\as(\mu_R)\over 4\pi}\left(1 - {\as(\mu_R)\over 4\pi}{\beta_0 \over \epsilon}
  +
{\mathcal{O}}(\as^2(\mu_R))\right) .
\label{eq:asren}
\end{equation}
The factor $S_\epsilon = \exp ( \epsilon \{\ln(4\pi)-\gamma_{\rm e}\} )$,
where $\gamma_{\rm e}$ denotes the Euler-Mascheroni constant,
is an artifact of dimensional
regularization~\cite{'tHooft:1972fi,Bollini:1972ui,Ashmore:1972uj,Cicuta:1972jf}
kept out of the coefficient functions and anomalous dimensions
in the $\MSbar$ scheme.
Also, we have
\begin{equation}
  \label{eq:beta0}
  \beta_0 = {11 \over 3} C_A - {4 \over 3} T_R n_f\, ,
\end{equation}
which is the ${\cal O}(\as^2)$ coefficient of the QCD $\beta$-function,
$C_A=3,~T_R=1/2$ are the QCD color factors,
$n_f$ denotes the number of light fermion flavors and $\mu_R$
stands for the renormalization scale. To simplify our expressions
we will work with $\mu_R^2=Q^2$. If desired, the explicit
dependence on the arbitrary renormalization scale can be easily restored.

To completely specify our observables, we have to clarify how the
bare partonic cross-sections are defined (and calculated).
This is done next in Section~\ref{sec:method}.

%
%
\section{The method}
\label{sec:method}
\subsection{General considerations}
\label{sec:generals}

Similarly to the DIS case, the bare
partonic differential distribution ${\cal F}$ for the process
Eq.~(\ref{eq:epem}) can be written as a product of a leptonic and
hadronic tensors. The hadronic tensor $W_{\mu\nu}$ is proportional
to the amplitude squared of the decaying vector boson,
and depends in particular on the coupling of the latter to the
fermion current. Throughout this paper we will only consider the
case of massless quarks\footnote{Mass effects are known
analytically through NLO~\cite{Nason:1993xx} and, based on a
numerical study, through NNLO~\cite{Nason:1998ug,Nason:1999zj}.}.
In this case, as was detailed in Ref.~\cite{Rijken:1996ns}, the
calculation of the inclusive QCD corrections to the corresponding
coefficient functions is independent of the vector boson
being vector (V) or axial-vector (A) type.
For its evaluation it is therefore sufficient to consider the
decay of a vector boson $V$ that couples to the fermion current as
$\gamma^\mu(1-\gamma^5)$.

The bare hadronic tensor $W_{\mu\nu}$ depends on two momenta: the
one of the decaying vector boson $q$ and the momentum of the
observed parton $p$. Following the usual arguments for Lorentz
and CP~invariance one can show that $W_{\mu\nu}$ can be parameterized
precisely with the three fragmentation functions ${\cal F}_{T,L,A}$
introduced in Eq.~(\ref{eq:mass-factorization}).
For the V-A couplings mentioned above
(see~\cite{Rijken:1996ns} for the general case),
these three functions can be expressed as projections of the
hadronic tensor $W_{\mu\nu}$ (with $d=4-2\epsilon$):
\begin{eqnarray}
\label{eq:Fproj} {\cal F}_{T}(z,\epsilon) &=& {1\over d-2} \left(
-2{p \cdot q\over q^2} W_\mu^{\,\,\mu} - {2\over p \cdot q} p^\mu
p^\nu W_{\mu\nu} \right)\, ,
\nonumber\\
{\cal F}_{L}(z,\epsilon) &=& {1\over p \cdot q} p^\mu p^\nu
W_{\mu\nu}\, ,
\nonumber\\
{\cal F}_{A}(z,\epsilon) &=& -{1\over q^2}{2\over (d-2)(d-3)}\,
i\, \epsilon^{\mu\nu\alpha\beta}p_\alpha q_\beta W_{\mu\nu}\, .
\end{eqnarray}

Our goal in this paper is to calculate the fragmentation functions
${\cal F}_{T,L,A}$ including the coefficient functions of order $\as^2$.
In fact, we even obtain the terms up to $\as^2\epsilon^2$
in the mass-factorization, some of which we have used to check
the analytical continuation of the space-like DIS results to the
time-like region~\cite{Mitov:2006ic}.
Terms of order $\as^2\epsilon^2$ would also be needed in a
future evaluation of the order $\as^4$ corrections to ${\cal F}_{T,L,A}$
(see also Eqs.~(\ref{eq:F0})--(\ref{eq:FAn2}) for more details on that point).

Next, we explain how to construct the contributions of order
$\as^k$ to the hadronic tensor $W_{\mu\nu}$. This tensor contains
the contributions from all diagrams for the processes,
\begin{equation}
\label{eq:process} V(q) \to h(p) + h_1(p_1) + \dots + h_c(p_c)\, ,
\quad\quad c=1\dots k+1\, ,
\end{equation}
of the decay of the vector boson $V$ with momentum $q$ to a set of
particles $h, h_1, \dots, h_{c}$. The different final states
(labeled by the number $c$ of unobserved particles in that state)
represent the contributions from the various physical cuts to the
single-particle inclusive observable. After the Feynman diagrams
contributing to each particular physical cut have been constructed
and appropriately added, one has to perform the required virtual
and/or (real) phase space integrations.

The approach we pursue for the evaluation of the required Feynman
integrals differs significantly from the calculation
in~\cite{Rijken:1996vr,Rijken:1996ns,Rijken:1996np}.
Our approach is based on the application of algebraic relations based on
integration by parts
(IBP)~\cite{'tHooft:1972fi,Tkachov:1981wb,Chetyrkin:1981qh,Tkachov:1984xk}
to cross-sections and it is performed {\it directly} in Mellin
$N$-space~\cite{Mitov:2005ps}.

Our starting point is Eq.~(\ref{eq:Fproj}); in the following we will
use ${\cal F}$ to denote any one of the three bare cross-sections
${\cal F}_{T,L,A}$. Each one of the functions ${\cal F}$ has a
perturbative expansion in terms of the {\it bare}\footnote{
We find it more convenient to present results in terms
of the bare coupling instead of the renormalized one, the reason
being that we work with massless on-shell partons. Thus we only
need UV renormalization, i.e. the one related to the coupling
Eq.~(\ref{eq:asren}).}
strong coupling,
\begin{equation}
\label{eq:expansion-F}
{\cal F}(n,\epsilon) =
\sum_{k=0}^\infty \left( {\alpha_s \over 4\pi} S_\epsilon \right)^k
{\cal F}^{(k)}(n,\epsilon)\, .
\end{equation}

Throughout this paper, the letter $n$ will be reserved for the
Mellin variable of any function of $z$ defined through
\begin{equation}
\label{eq:Mellin-n}
f(n) = \int_0^1 dz\,~ z^n f(z)\, .
\end{equation}
with $n\geq 0$. In particular, the total integral of a function
corresponds to $n=0$. This definition of the Mellin variable is
the most natural choice for the calculational procedures detailed
in the following. We will, however, present the final results for
the corresponding finite partonic cross-sections $\hat{\cal F}$ in
terms of the conventional Mellin variable $N$ defined through
\begin{equation}
\label{eq:Mellin-N}
{\cal F}(N,\epsilon) = \int_0^1 dz\,~ z^{N-1} {\cal F}(z,\epsilon)\, .
\end{equation}
In view of the additional factor of $z$ in the partonic equivalent
of Eq.~(\ref{eq:sigmatot}) the relation between the two variables
is
\begin{equation}
n=N-2\, .
\label{eq:n-N}
\end{equation}

Each function ${\cal F}^{(k)}(n,\epsilon)$ with $k\geq 0$ contains
contributions from a number of terms, corresponding to the
different physical cuts of the process $V \to h+X$ at order
$\as^k$:
\begin{equation}
\label{eq:cuts}
{\cal F}^{(k)}(n,\epsilon) = \sum_{c=1}^{k+1} {\cal
F}^{(k)}_{(c)}(n,\epsilon)\, .
\end{equation}

The functions ${\cal F}^{(k)}_{(c)}(n,\epsilon)$ contain the full
contributions from the process $V\to h +X$ at order $\as^k$ where
the inclusive final state $X$ contains $c$ unresolved partons. As
described in Ref.~\cite{Mitov:2005ps}, the construction of the
functions ${\cal F}^{(k)}_{(c)}(n,\epsilon)$ consists of the
following steps:
\begin{enumerate}
\item One constructs all contributing Feynman diagrams.
The integrations over the virtual momenta $\displaystyle \prod_i
{d^dk_i\over (2\pi)^d}$ are assumed implicit in the diagrams.

\item One constructs the corresponding contribution to the tensor
$W_{\mu\nu}$ by adding all relevant amplitudes squared, and with
the appropriate symmetry factors included.

\item The above result is contracted with the appropriate tensor
constructed from the $d$-dimensional metric and the momenta $p$ and $q$
as follows from Eq.~(\ref{eq:Fproj}).

\item The Lorentz scalar constructed this way is
integrated over the full phase-space of the $c+1$ partons (i.e.
one also integrates over the full phase-space of the `observed'
parton). The measure for this integration is:
\begin{eqnarray}
\label{eq:ps}
d\Phi &=& \left( 2p \cdot q \right)^n ~ (2\pi)^{d}\delta(q-p-p_1-\dots-p_c)
\prod_{i=0}^c {d^dp_i\over (2\pi)^{d-1}} \delta(p_i^2) .
\end{eqnarray}
We have defined $p_0\equiv p$, and we have set $Q^2=1$ for
simplicity. The exact dependence on $Q^2$ can be easily restored
on dimensional grounds. The origin of the `Mellin propagator'
$2p \cdot q$ is explained at the end of this Subsection.

\item One applies the IBP identities to reduce each term
(generally containing integrations over both real and virtual
momenta) to a linear combination of a small number of independent
`master' integrals. As a rule, at order $\as^k$ and for each
particular cut $c$, one needs to construct and solve more than one
IBP reduction; the number of the required reductions corresponds
to the number of independent topologies for each set $(k,c)$. The
$\delta$-functions from the real-phase space are dealt with along
the lines of Ref.~\cite{Anastasiou:2002yz}, while the Mellin
propagator is treated along the lines of Ref.~\cite{Mitov:2005ps}.

\item Each master integral is a function of the Mellin variable $n$.
Its $n$-dependence can be completely extracted with the help
of the difference equation the masters satisfy. The difference
equations are obtained from the solutions to the IBP reduction
(see also \cite{Vermaseren:2005qc,Moch:2002sn} for
related discussions in the DIS case).

\item One has to supply appropriate initial conditions for
specifying the solutions of the difference equations. The most
suitable choice is to evaluate the value of the masters at $n=0$.
This choice corresponds to the total integral of each master over
$z$. Therefore the initial conditions are pure,
$\epsilon$-dependent numbers.

\item Following~\cite{Mitov:2005ps}, we `partial fraction' by
performing an additional summation over $n$ of the terms
containing a propagator of the type $\sim 1/(1-2p \cdot q)$.
This propagator is not linearly independent from the `Mellin propagator'
$2p \cdot q$ as it merely shifts the effective $n$ in
complete analogy to the DIS case,
see e.g. Refs.~\cite{Moch:1999eb,Moch:2002sn}.

\end{enumerate}
For the evaluation of the transverse and the longitudinal
functions it is sufficient to take the matrix $\gamma^5$ as
anti-commuting in $d$-dimensions. Special care is, however, needed
for the evaluation of the asymmetric contribution ${\cal F}_{A,\, {\rm ns}}$.

For the evaluation of ${\cal F}_{A,\, {\rm ns}}$ we follow the
prescription of Larin~\cite{Larin:1993tq}. The details about the
implementation can be found e.g.
in~\cite{Moch:1999eb,Moch:2004pa}. In short, there are two
important features: First, one replaces the axial-vector coupling
$\gamma^\mu\gamma^5$ with the $d$-dimensional completely
antisymmetric tensor $\epsilon^{\mu \rho \sigma \tau} \gamma_\rho
\gamma_\sigma \gamma_\tau$ and then uses its contraction
properties with the second $\epsilon$-tensor appearing in
Eq.~(\ref{eq:Fproj}) to reduce it to combinations of the
$d$-dimensional metric tensor. Second, one has to multiply the
resulting expression with additional renormalization constants
which restore the axial Ward identity in the $\MSbar$-scheme.
These constants have an expansion in the renormalized coupling
$\as$ and in powers of $\epsilon$. They have been computed to
three-loops in~\cite{Larin:1991tj} and take the following form:
\begin{eqnarray}
\label{eq:Z5ZA}
Z_5 &=&
    1
  + {\alpha_s(\mu_R) \over 4\pi} \cf
    \biggl\{ - 4 - 10 \epsilon  + (-22 + 2\z2 ) \epsilon^2
      + {\cal O}(\epsilon^3) \biggr\}
\nonumber\\
&& + \left({\alpha_s(\mu_R) \over 4\pi}\right)^2
    \biggl\{ 22 \cf^2 - {107 \over 9} \ca \cf
      +{ 2 \over 9} \nf \cf
\nonumber\\
&&\mbox{} + \left( (132 - 48 \z3) \cf^2
  + \left( - {7229 \over 54} + 48\z3 \right) \ca \cf
      + {331\over 27} \nf \cf \right) \epsilon + {\cal O}(\epsilon^2) \biggr\}
  + {\cal O}(\as^3)
\nonumber\\
Z_A &=&
    1
   + \left({\alpha_s(\mu_R) \over 4\pi}\right)^2 {1\over \epsilon}
     \left\{ {22\over 3} \ca \cf - {4\over 3} \nf \cf \right\}
   + {\cal O}(\as^3) \, .
\end{eqnarray}

We would like to conclude this Subsection with a comment on the
origin of the `Mellin propagator' $P_M = 2p \cdot q$ appearing in
Eq.~(\ref{eq:ps}). As was detailed in \cite{Mitov:2005ps}, to
construct the bare distribution ${\cal F}(z,\epsilon)$ in $z$-space, one
has to integrate over the full phase space of all final states
particle and insert the additional factor $\delta(z-2p \cdot q)$. If one
Mellin-transforms this expression before the required phase-space
and virtual integrations are performed, one gets schematically
(see also Eqs.~(\ref{eq:def-z}), (\ref{eq:ps})):
\begin{eqnarray}
\label{eq:nproj}
{\cal F}(n,e) &=& (\dots)~\times~ \int_0^1
dz~z^n~\delta(z-2p \cdot q)
\nonumber\\
&=& (\dots)~\times~ (2p \cdot q)^n
\, ,
\end{eqnarray}
i.e. the factor of the `Mellin propagator' raised to a
{\it symbolic} power $n$ that appears in Eq.~(\ref{eq:ps}). The factor
$(\dots)$ stands for the various propagators (including possibly
additional powers of $P_M$), the measures for the real and/or
virtual integrations, etc., but contains no dependence on $z$ or $n$.
This procedure is completely analogous but more general than
the corresponding DIS
case~\cite{Kazakov:1987jk,Moch:1999eb,Moch:2004pa,Vogt:2004mw,Vermaseren:2005qc},
which relies on the OPE and the method
of projection to directly expand propagators in powers of $(2 p \cdot q/q^2)^n$.
There Eq.~(\ref{eq:nproj}) is effectively realized by
mapping any Feynman diagram to Mellin moments with the help of a
suitable projection operator~\cite{Gorishnii:1983su,Gorishnii:1987gn}.

In the following we will present the specifics of the
implementation of the above procedure for the evaluation of the
contributions at orders $\as$ and $\as^2$.

\subsection{Order ${\cal O}(\as)$}
\label{sec:order-as^1}

Here we discuss both the derivation and the results for all
independent contributions at order ${\cal O}(\as)$.
This order is the NLO result for the transverse and the asymmetric
contributions, but the LO result for the longitudinal function,
since it vanishes at order ${\cal O}(\as^0)$.

At order ${\cal O}(\as)$ in the perturbative expansion, one has to
consider only two cuts: one corresponding to real gluon emission
(where the final state is $(q\bar{q}g)$) and the one with the
virtual correction to the $Vq\bar{q}$ vertex (with the final state
being $(q\bar{q})$). The evaluation of these two contributions is
fairly different. We discuss first the contribution from the
virtual corrections.

From the kinematics of the tree-level decay process $V\to
q\bar{q}$ it is clear that the contribution of the purely virtual
corrections to the functions ${\cal F}$ are of the type
$const(\epsilon)\delta(1-z)$ in $z$-space which corresponds to a
$n$-independent constant in $n$-space. Therefore, the contribution
from this cut is completely determined by (twice) the real part of
the time-like
form-factor~\cite{vanNeerven:1985xr,Matsuura:1988sm,Moch:2005id}.

All non-trivial $n$-dependence at that order comes from the
real-emission diagrams. The corresponding diagrams can be found in
Ref.~\cite{Rijken:1996ns} and we do not repeat them here.
As was outlined in the previous Subsection, for the
evaluation of all contributions ($T,L,A$) for both a quark and a
gluon, one needs to construct a single topology consisting of five
`propagators' (see also~\cite{Mitov:2005ps}). One of the arguments,
of course, corresponds to the Mellin propagator $2p \cdot q$. We
performed the required IBP reductions and obtained a single
$n$-dependent master integral, which can be found e.g.
in~\cite{Mitov:2005ps}.

To perform the reductions resulting from the IBP identities, we
have used the program {\sc AIR}~\cite{Anastasiou:2004vj} which is
an implementation of the so-called Laporta algorithm~\cite{vanRitbergen:1999fi,%
Laporta:2000dc,Laporta:2001dd,Schroder:2002re} in {\sc Maple}.
{\sc AIR} contains also a routine which conveniently and
automatically maps the constructed diagrams into the master
integrals of the reductions (a single master at this order).
Following this simple procedure, one can map the whole problem of
the evaluation of any one of the functions ${\cal F}$ at order
$\as^1$ to the single master multiplied by a rational function of
$n$ and $\epsilon$. As discussed above there is another type of
contributions containing a propagator that is not linearly
independent of the Mellin propagator $P_M=2p \cdot q$, i.e.
contributions of the form:
\begin{eqnarray}
\label{eq:parfrac}
(\dots) {P_M^n\over 1-P_M} = \sum_{k=0}^\infty (\dots) P_M^{n+k}
\end{eqnarray}
where dots stand for any other, linearly independent propagator
and the integration measure over the real momenta. Since the term
on the right hand side of Eq.~(\ref{eq:parfrac}) is of the usual
form, the results from the solutions to the IBP reduction can be
applied (with $n$ replaced by $n+k$) and then summed over $k$. To
illustrate that point, we present the corresponding term of the
order $\as^1$ contribution to the function ${\cal F}^{T}_q$:
\begin{eqnarray}
&&C_F~\sum_{k=n}^\infty~
\epsilon(1-\epsilon)\left(4\epsilon^3-4\epsilon^2k-54\epsilon^2+31\epsilon k+74\epsilon+\epsilon
k^2-4k^2-20k-24\right)
\nonumber\\
&&\mbox{} \times
{ \Gamma(-\epsilon)^2\Gamma(k+1-2\epsilon) \over
\Gamma(2-2\epsilon)\Gamma(4-3\epsilon+k) }
\, ,
\end{eqnarray}
which can be easily summed up in terms of $\Gamma$-functions. This
way, one arrives at a very compact result
for the bare $(T,L,A)$ partonic cross-sections at order $\as$
valid to all orders in $\epsilon$.
The explicit results, expanded to sufficient
powers in $\epsilon$, can be found in Section~\ref{sec:results}.

%
%
\subsection{Order $O(\as^2)$}
\label{sec:order-as^2}

At this order one has to consider three different cuts:
double-virtual corrections with final state ($q,\bar{q}$),
one-loop virtual corrections to the final state ($q,\bar{q},g$)
and, finally, the cut with double real emission. The latter
consists of the following final states: $(q,\bar{q},g,g)$;
$(q,\bar{q},q,\bar{q})$ and $(q,\bar{q},q',\bar{q}')$. Depending
on the gauge choice for the polarization of the external gluons
one may also have to consider external ghosts. Again the
corresponding diagrams can be found in Ref.~\cite{Rijken:1996ns}
and we are only interested in the evaluation of the diagrams with
real emissions since the diagrams with two-loop virtual
corrections produce only constant terms in Mellin $n$-space. These
purely virtual contributions can be obtained from the one- and
two-loop time-like
form-factor~\cite{vanNeerven:1985xr,Matsuura:1988sm,Moch:2005id}.

To cover all possible diagrams for the evaluation of both quark
and gluon production we construct seven topologies for the double
real emission cut, and five topologies for the real-virtual cut.
After symmetry considerations, we arrive at a total of six
real-real and five different real-virtual masters. These
$n$-dependent masters satisfy difference equations that can be
read off the completed reductions. For completeness, we have
presented both the definitions of the masters and the difference
equations they satisfy in Appendix A.

As can be seen there, the structure of the equations is simple.
The simplest masters decouple and satisfy homogeneous equations,
while the more complicated ones satisfy first order difference
equations with the non-homogeneous part comprised by simpler,
explicitly known masters. Such first order equations can be easily
solved in closed form to all orders in $n$ and $\epsilon$, see
e.g. Ref.~\cite{Moch:1999eb}. If we pursue this approach, the most
complicated solutions we encounter are Appel functions of unit
arguments or hypergeometric functions of unit argument.

However, we decided to follow a different path.
One can also obtain the solutions of the difference equations after an
expansion in powers of $\epsilon$ using the methods
of symbolic summation and the packages
{\sc Summer}~\cite{Vermaseren:1998uu} and {\sc XSummer}~\cite{Moch:2005uc} in
{\sc Form}~\cite{Vermaseren:2000nd}. Then it is very easy to solve the
equations this way given that previously even three-loop master integrals
have been computed~\cite{Moch:2004pa,Moch:2005id,Moch:2005uc}.
In this approach, one is required to supply the initial conditions
(for $n=0$) beforehand. This is to be contrasted with the all-order
in $\epsilon$ calculation where the initial condition factorizes
completely.

Only one master, $R_2(n)$ deserves special consideration. It
formally satisfies a second order difference equation as can be
seen from Eq.~(\ref{eq:rm2-n}). A closer inspection, however,
reveals that it is a second order difference equation of defined
parity (thus a $n-1$ term is absent). Therefore one can write this
equation as a first order difference equation for a `new variable'
$n' = n/2$. In doing so we could solve the resulting difference
equation in terms of $_7F_6$-type functions of unit argument.
These functions contain half-integer (and $n/2$) parameters and
are not simple to expand in $\epsilon$. If one, however, solves
this master as an expansion in powers of $\epsilon$ no particular
complications arise besides the fact that one has to supply two
initial conditions (for $n=0$ and for $n=1$) for any second order
equation.

Next we address the calculation of the initial conditions for the
masters. Our first step is to perform new IBP reductions in each
of the real-real and real-virtual topologies setting $n=0$ from
the very beginning. From this fixed order reductions we obtain
relations between the initial conditions for the masters integrals.
In the end, we find that a total of seven are independent.
Most of these integrals are very easy to compute directly to all orders
in $\epsilon$.
The most involved is the initial condition $R_6(0)$
given in Eq.~(\ref{eq:rm6}).
For its evaluation (up to weight 6 in values of the Riemann zeta-function)
we have used the approach of Ref.~\cite{DeRidder:2003bm}
based on the optical theorem and the
results of Ref.~\cite{Kazakov:1984km} for the higher order
$\epsilon$ terms of the non-planar three-loop two-point function.
With the help of these additional fixed-$n$ runs we can derive a relation (see
Eq.~(\ref{eq:rm2-1})) between the values of the real master $R_2$
at $n=0$ and $n=1$ as mentioned above.

The last point that deserves special attention are terms $\sim 1/(1-2p \cdot q)$.
`Partial fractioning' of these leads to an additional sum as explained
above, cf. Eq.~(\ref{eq:parfrac}).
In performing the required symbolic summation we employ the following strategy.
We explicitly separate sums from $1$ to $\infty$ over $n$-independent terms,
as we systematically ignore constant terms from the
evaluation of the diagrams with real emissions as well as the
constant terms from the purely virtual corrections.
After we complete our evaluation we can restore these constant terms from
the requirement that the total cross-section is reproduced
(see Eq.~(\ref{eq:sigmatot})).
We wish to emphasize, though, that this procedure is simply done to economize
on the necessary algebra and by no means represents any principal drawback
of our approach.

The above comment applies to the transverse and asymmetric
partonic cross-sections. Since the longitudinal cross-sections do
not contain any constant terms, this procedure uniquely fixes the
missing constant contributions in the transverse functions. It
also provides the coefficient of the $\delta$-function
contributions to the asymmetric functions. To that end, one uses
the fact that the difference of the transverse and asymmetric
functions when expressed in $z$-space, does not contain
$\delta$-functions and singular +-distributions. Equivalently,
the difference of these functions vanishes in the `soft' limit
$N\to\infty$.

%
%
\section{Results}
\label{sec:results}

Let us now present the results of the calculation. As explained
above, mass factorization (or collinear renormalization) predicts
a specific structure for the final result, which we write out up
to second order in the strong coupling $a_{\rm s} = \as/(4\pi)$.
Following the conventions from Eq.~(\ref{eq:expansion-F}), we
present the general structure of the result expanded in $\epsilon$
directly in Mellin $N$-space. According to
Eq.~(\ref{eq:Mellin-N}), the Mellin moments of the splitting
functions are defined by
\begin{eqnarray}
  \label{eq:defP}
  P(N) \,=\, \int_0^1 dz\,~ z^{N-1} P(z) \,=\, -\, \gamma(N)\, ,
\end{eqnarray}
and we note the (conventional) sign for the relation of the
splitting functions to the anomalous dimensions. In the following,
all products are to be evaluated employing the algebra for harmonic
sums~\cite{Vermaseren:1998uu,Moch:1999eb}.

The zeroth-order contributions, with ${\cal F}^{(0)}_{T,\rm q}$ being
suitably normalized, read
\begin{equation}
\label{eq:F0}
  {\cal F}_{T,\rm q}^{(0)} \: = \: c_{T,\rm q}^{(0)} \: = \: 1
   \:\: , \qquad
  {\cal F}_{T,\rm g}^{(0)} \: = \: {\cal F}_{L,\rm q}^{(0)}
  \: = \: {\cal F}_{L,\rm g}^{(0)} \: = \: 0
   \:\: , \qquad
  {\cal F}_{A,\rm q}^{(0)} \: = \: c_{A,\rm q}^{(0)} \: = \: 1
  \:\: .
\end{equation}
Note that not all functions are independent; on general grounds
one can show that \cite{Rijken:1996ns}:
\begin{eqnarray}
\label{eq:Fqgen}
{\cal F}_{T,\, \rm q} &=& +{\cal F}_{T,\, \rm \bar{q}}\, ,
\nonumber \\
{\cal F}_{A,\, \rm q} &=& -{\cal F}_{A,\, \rm \bar{q}}\, ,
\nonumber \\
{\cal F}_{A,\, \rm g} &=&  0\, .
\end{eqnarray}

For the applications of the present study we present the
amplitudes at the first order in $\as$ up to terms of the order
$\epsilon^2$, yielding for ${\cal F}_{T}$
\begin{eqnarray}
\label{eq:FT1q}
  {\cal F}_{T, \rm q}^{(1)} & = &
  - {1 \over \epsilon}\, P_{\,\rm qq}^{\,(0)}
  \: + \: c_{T,\rm q}^{(1)} \: + \: \epsilon\, a_{T,\rm q}^{(1)}
  \: + \: \epsilon^2 b_{T,\rm q}^{(1)} \, ,
\\
\label{eq:FT1g}
  {\cal F}_{T, \rm g}^{(1)} & = &
  - {2 \over \epsilon}\, P_{\,\rm gq}^{\,(0)}
  \: + \: c_{T,\rm g}^{(1)} \: + \: \epsilon\, a_{T,\rm g}^{(1)}
  \: + \: \epsilon^2 b_{T,\rm g}^{(1)} \, ,
\end{eqnarray}
for the longitudinal ${\cal F}_{L}$
\begin{eqnarray}
\label{eq:FL1q}
  {\cal F}_{L, \rm q}^{(1)}  & = &  c_{L,\rm q}^{(1)}
  \: + \: \epsilon\, a_{L,\rm q}^{(1)} \: + \: \epsilon^2 b_{L,\rm q}^{(1)}
  \:\: , \\[0.5mm]
\label{eq:FL1g}
  {\cal F}_{L, \rm g}^{(1)}  & = &  c_{L,\rm g}^{(1)}
  \: + \: \epsilon\, a_{L,\rm g}^{(1)} \: + \: \epsilon^2 b_{L,\rm g}^{(1)}
  \:\: ,
\end{eqnarray}
and for the asymmetric ${\cal F}_{A}$
\begin{eqnarray}
\label{eq:FA1ns}
  {\cal F}_{A, \rm ns}^{(1)} & = &
  - {1 \over \epsilon}\, P_{\,\rm qq}^{\,(0)}
  \: + \: c_{A,\, \rm q}^{(1)} \: + \: \epsilon\, a_{A,\, \rm q}^{(1)}
  \: + \: \epsilon^2 b_{A,\, \rm q}^{(1)}
  \:\: .
\end{eqnarray}

Correspondingly, the $\as^2$ contributions where the non-singlet
and singlet quark amplitudes differ for the first time, are
required up to order $\epsilon$. These quantities are given by
\begin{eqnarray}
\label{eq:FTn2}
  {\cal F}_{T,\rm ns}^{(2)} & \!=\! &
  {1 \over 2\epsilon^2}\, \bigg\{ P_{\,\rm qq}^{\,(0)} \left( P_{\,\rm qq}^{\,(0)} + \beta_0 \right) \bigg\}
  \: - \: {1 \over 2\epsilon}\, \left\{ P_{\,\rm ns}^{\,(1)+}
    + 2\, c_{T,\rm q}^{(1)}\, P_{\,\rm qq}^{\,(0)} \right\}
\nonumber \\ & & \mbox{}
  \: + \: {c_{T,\rm ns}^{(2)}} - a_{T,\rm q}^{(1)}\, P_{\,\rm qq}^{\,(0)}
  \: + \: \epsilon\, \bigg\{ {a_{T,\rm ns}^{(2)}} - b_{T,\rm q}^{(1)}\, P_{\,\rm qq}^{\,(0)} \bigg\} \:\: ,
  \\[2mm]
\label{eq:FTps2}
  {\cal F}_{T,\rm ps}^{(2)} & \!=\! & {1 \over 2\epsilon^2}\,
  \bigg\{ P_{\,\rm qg}^{\,(0)} P_{\,\rm gq}^{\,(0)} \bigg\}
  \: - \: {1 \over 2\epsilon}\,
  \left\{ P_{\,\rm qq}^{\,(1) {\rm s}} + c_{T,\rm g}^{(1)}\, P_{\,\rm qg}^{\,(0)} \right\}
\nonumber \\ & & \mbox{}
  \: + \: {c_{T,\rm ps}^{(2)}} - {1 \over 2}\, a_{T,\rm g}^{(1)}\, P_{\,\rm qg}^{\,(0)}
  \: + \: {1 \over 2}\, \epsilon\, \bigg\{ 2 {a_{T,\rm ps}^{(2)}}  - b_{T,\rm g}^{(1)}\, P_{\,\rm qg}^{\,(0)} \bigg\} \:\: ,
  \\[2mm]
\label{eq:FTg2}
  {\cal F}_{T,\rm g}^{(2)} & \!=\! & {1 \over \epsilon^2}\,
  \bigg\{ P_{\,\rm gq}^{\,(0)} \left( P_{\,\rm qq}^{\,(0)} + P_{\,\rm gg}^{\,(0)} + \beta_0 \right) \bigg\}
  \: + \: {1 \over \epsilon}\, \bigg\{ P_{\,\rm gq}^{\,(1)} + 2\, c_{T,\rm q}^{(1)}\, P_{\,\rm gq}^{\,(0)}
   + c_{T,\rm g}^{(1)}\, P_{\,\rm gg}^{\,(0)} \bigg\}
\nonumber \\ & & \mbox{}
  \: + \: c_{T,\rm g}^{(2)} - 2\, a_{T,\rm q}^{(1)}\, P_{\,\rm gq}^{\,(0)}
   - a_{T,\rm g}^{(1)}\, P_{\,\rm gg}^{\,(0)}
  \: + \:  \epsilon\, \bigg\{ a_{T,\rm g}^{(2)} - 2\, b_{T,\rm q}^{(1)}\, P_{\,\rm gq}^{\,(0)}
   - b_{T,\rm g}^{(1)}\, P_{\,\rm gg}^{\,(0)} \bigg\} \:\: ,
\end{eqnarray}
\begin{eqnarray}
\label{eq:FLn2}
  {\cal F}_{L,\rm ns}^{(2)} & \!=\! & - {1 \over \epsilon}\,
  \bigg\{ c_{L,\rm q}^{(1)}\, P_{\,\rm qq}^{\,(0)} \bigg\}
  \: + \: c_{L,\rm ns}^{(2)} - a_{L,\rm q}^{(1)}\, P_{\,\rm qq}^{\,(0)}
  \: + \: \epsilon\, \bigg\{ a_{L,\rm ns}^{(2)} - b_{L,\rm q}^{(1)}\, P_{\,\rm qq}^{\,(0)} \bigg\} \:\; ,
  \\[2mm]
\label{eq:FLps2}
  {\cal F}_{L,\rm ps}^{(2)} & \!=\! & - {1 \over \epsilon}\,
  \bigg\{ c_{L,\rm g}^{(1)}\, P_{\,\rm qg}^{\,(0)}  \bigg\}
  \: + \: {c_{L,\rm ps}^{(2)}} - {1 \over 2}\, a_{L,\rm g}^{(1)}\, P_{\,\rm qg}^{\,(0)}
  \: + \: {1 \over 2}\, \epsilon\, \bigg\{ 2 {a_{L,\rm ps}^{(2)}}
     - b_{L,\rm g}^{(1)}\, P_{\,\rm qg}^{\,(0)} \bigg\} \:\: ,
  \\[2mm]
\label{eq:FLg2}
  {\cal F}_{L,\rm g}^{(2)} & \!=\! & {1 \over \epsilon}\,
  \bigg\{ 2\, c_{L,\rm q}^{(1)}\, P_{\,\rm gq}^{\,(0)} + c_{L,\rm g}^{(1)}\, P_{\,\rm gg}^{\,(0)}\bigg\}
  \: + \: c_{L,\rm g}^{(2)} - 2\, a_{L,\rm q}^{(1)}\, P_{\,\rm gq}^{\,(0)}
   - a_{L,\rm g}^{(1)}\, P_{\,\rm gg}^{\,(0)}
\nonumber \\ & & \mbox{}
  \: + \:  \epsilon\, \bigg\{ a_{L,\rm g}^{(2)} - 2\, b_{L,\rm q}^{(1)}\, P_{\,\rm gq}^{\,(0)}
   - b_{L,\rm g}^{(1)}\, P_{\,\rm gg}^{\,(0)} \bigg\} \:\: .
\end{eqnarray}
\begin{eqnarray}
\label{eq:FAn2}
  {\cal F}_{A,\rm ns}^{(2)} & \!=\! &
  {1 \over 2\epsilon^2}\, \bigg\{ P_{\,\rm qq}^{\,(0)} \left( P_{\,\rm qq}^{\,(0)} + \beta_0 \right) \bigg\}
  \: - \: {1 \over 2\epsilon}\, \left\{ P_{\,\rm ns}^{\,(1)-}
    + 2\, c_{A,\rm q}^{(1)}\, P_{\,\rm qq}^{\,(0)} \right\}
\nonumber \\ & & \mbox{}
  \: + \: {c_{A,\rm ns}^{(2)}} - a_{A,\rm q}^{(1)}\, P_{\,\rm qq}^{\,(0)}
  \: + \: \epsilon\, \bigg\{ {a_{A,\rm ns}^{(2)}} - b_{A,\rm q}^{(1)}\, P_{\,\rm qq}^{\,(0)} \bigg\} \:\: .
\end{eqnarray}
The results for the splitting and coefficient functions at order ${\cal O}(\as)$
and ${\cal O}(\as^2)$ are given both in $N$-space (Appendix B) and
$x$-space (Appendix C).
The precise definition of the various splitting functions can be
found in Eqs.~(30)--(34) in \cite{Moch:1999eb},
see also~\cite{Ellis:1991qj}
for a more detailed discussion on that point.

Several comments are in order:
First of all, as we have performed the calculation in Mellin space, all
$x$-dependence is recovered from the $N$-space results by an inverse Mellin
transformation, which expresses these functions in terms of harmonic
polylogarithms~\cite{Remiddi:1999ew}.
The inverse Mellin transformation exploits an isomorphism between the
set of harmonic sums for even or odd $N$ and the set of harmonic
polylogarithms (see also Appendix B).
The algebraic procedure~\cite{Moch:1999eb,Remiddi:1999ew}
is based on the fact that harmonic sums occur as coefficients of the
Taylor expansion of harmonic polylogarithms.

Our results for the finite terms in $\epsilon$ agree with the ones
in~\cite{Rijken:1996vr,Rijken:1996ns,Rijken:1996np}. We have found
several misprints in these references and would like to take the
opportunity to point out these typos in the original manuscript
of Ref.~\cite{Rijken:1996ns} (employing the notation of the original
reference).
In Eq.~(A.6) of Ref.~\cite{Rijken:1996ns} for
${\bar c}_{T,q}^{\rm{NS},(2),\rm{nid}}\bigl|_H$ there should be a
replacement of the term
\begin{equation}
  \label{eq:RvNtypo1}
\cf^2 (1+z) ( \dots - 3 \ln^2 z \dots ) \quad \longrightarrow
\quad \cf^2 (1+z) ( \dots - 3 \ln^3 z \dots ) \, ,
\end{equation}
in Eq.~(A.8) of Ref.~\cite{Rijken:1996ns} for ${\bar
c}_{T,q}^{\rm{NS},(2),\rm{id}}$ of the term
\begin{equation}
  \label{eq:RvNtypo2}
(\cf^2-{1 \over 2} \ca\cf ) \left( {24 \over 5 z^2} + \dots
\right) \ln z \quad \longrightarrow \quad (\cf^2-{1 \over 2}
\ca\cf ) \left( {24 \over 5 z} + \dots \right) \ln z \, ,
\end{equation}
in Eq.~(A.10) of Ref.~\cite{Rijken:1996ns} for ${\bar
c}_{T,q}^{\rm{PS},(2)}$ of the term
\begin{equation}
  \label{eq:RvNtypobraket}
\cf T_f\left[ \dots + {11\over 6}\ln^3z +\dots\right. \quad
\longrightarrow \quad  \cf T_f\left[ \dots + \left. {11\over
6}\ln^3z \right) +\dots\right.,
\end{equation}
and in Eq.~(A.15) of Ref.~\cite{Rijken:1996ns} for ${\bar
c}_{L,q}^{\rm{NS},(2),\rm{id}}$ of the term
\begin{equation}
  \label{eq:RvNtypo3}
(\cf^2-{1 \over 2} \ca\cf ) \left( 32 S_{1,2}(1-z) + \dots \right)
\quad \longrightarrow \quad (\cf^2-{1 \over 2} \ca\cf ) \left( 32
S_{1,2}(-z) + \dots \right) \, .
\end{equation}
Finally, in Eq.~(17) of Ref.~\cite{Rijken:1996np} for ${\bf
C}_{A,q}^{\rm{NS},\rm{nid},(2)} - {\bf
C}_{T,q}^{\rm{NS},\rm{nid},(2)}$ one should replace the term
\begin{equation}
\cf^2 \left( - {24 \over 5 z^2} + \dots \right) \ln z \quad
\longrightarrow \quad \cf^2 \left( - {24 \over 5 z} + \dots
\right)\ln z \, .
\end{equation}

Beyond the coefficient functions at order ${\cal O}(\as^2)$ we have also
obtained the terms of higher order in the $\epsilon$ expansion, specifically
$a^{(1)},b^{(1)}$ and $a^{(2)}$.
In the non-singlet case, we have found them to agree with the predictions based
on the analytical continuation proposed in~\cite{Mitov:2006ic}.
However, as these expressions ($a^{(1)},b^{(1)}$ and $a^{(2)}$)
are particularly lengthy and have no direct physical application, we refrain from
writing them out explicitly here.

For future use and for uniformity of the notations, we present in
the Appendices B and C also the explicit expressions for the one-
and two-loop time-like splitting functions. Our calculations agree
with the known results \cite{Curci:1980uw,Furmanski:1980cm} (see
also \cite{Ellis:1991qj}). This statement, though, is subject to
one qualification. As we are considering in Eq.~(\ref{eq:epem})
only the decay of a vector boson $V$, we have no access to the
two-loop (time-like) splitting functions $P_{\,\rm qg}^{\,(1)}$
and $P_{\,\rm gg}^{\,(1)}$ with the set of Feynman diagrams
considered. To do so, we would actually be required to compute
also the decay of a (fictitious) classical scalar $\phi $ that
couples directly only to the gluon field via $\phi\,
G_{\mu\nu}^{\,a}G_a^{\,\mu\nu}$. In time-like kinematics, this
approach has been used for instance to derive $P_{\,\rm
gg}^{\,(1)}$ in \cite{Kosower:2003np}; see also
Refs.~\cite{Vogt:2004mw,Vermaseren:2005qc,Larin:1997wd} for the
analogous considerations in the space-like case. An inclusion of
the $\phi\, G_{\mu\nu}^{\,a}G_a^{\,\mu\nu}$ coupling is straight
forward and would allow for the determination of $P_{\,\rm
qg}^{\,(1)}$ and $P_{\,\rm gg}^{\,(1)}$ (or rather its Mellin
transform) to the desired two-loop accuracy.

%
%
\section{Summary}
\label{sec:summary}

We have calculated the ${\cal O}(\as^2)$ corrections to the
transverse, longitudinal and asymmetric fragmentation functions
for both quarks and gluons in semi-inclusive $e^+e^-$-annihilation
to hadrons. Our calculation confirms the results of Rijken and
van Neerven~\cite{Rijken:1996ns,Rijken:1996vr,Rijken:1996np} and
we have taken the opportunity to correct several typographical errors
in these papers (see also \cite{Mitov:2006ic}).
Our results constitute a strong check
on~\cite{Rijken:1996ns,Rijken:1996vr,Rijken:1996np}, in particular
since we have used a rather different technology and obtained
them directly in Mellin $N$-space following the proposal
in \cite{Mitov:2005ps}.
Thus, our calculation represents the first example of a
single-particle inclusive observable
beyond the well established DIS
framework~\cite{Kazakov:1987jk,Moch:1999eb,Moch:2004pa,Vogt:2004mw,Vermaseren:2005qc}
that is computed analytically in Mellin $N$-space.

The coefficient functions presented in this paper contain the NNLO
corrections to the transverse and asymmetric fragmentation
functions ${\cal F}_{T}$ and ${\cal F}_{A}$, and the NLO
corrections to the longitudinal ones, ${\cal F}_{L}$. After the
complete singlet three-loop time-like splitting functions become
available (the non-singlet case has recently been
reported~\cite{Mitov:2006ic}), one will be able to study light and
heavy quark fragmentation at
NNLO~\cite{Mele:1990cw,Melnikov:2004bm,Mitov:2004du}. This
important class of observables will allow precise extraction of
fragmentation distributions from LEP data and is yet another
motivation for the realization of the envisioned high-precision
Giga-$Z$ option of the future ILC. Due to the process independence
of the fragmentation distributions, they can be further applied to
other processes like hadro- and photo-production.

In addition, the present paper provides the means to thoroughly
study the relations between the time-like QCD corrections to
inclusive hadron production in $e^+e^-$ annihilation and their
space-like counterparts, the DIS structure functions. Our
calculational approach easily allows us to obtain higher powers in
$\epsilon$ of the bare partonic cross-sections at order ${\cal
O}(\as^2)$, which had not been computed before. Thus, we could
provide important cross checks on the procedure of
Ref.~\cite{Mitov:2006ic} based on an analytic continuation in $x$
between observables with space- and time-like kinematics. Of
course, the aforementioned higher terms in $\epsilon$ will also be
needed for a future evaluation of the QCD corrections to
$e^+e^-$-annihilation at order ${\cal O}(\as^3)$. Finally, one
particularly appealing feature of the $N$-space approach is the
small number of master integrals that have to be evaluated.
Moreover, with the boundary conditions in $N$-space being
kinematics independent, the corresponding integrals may also be of
relevance in other circumstances. As a matter of fact, some of
them had been considered before in a different
context~\cite{DeRidder:2003bm}. In the present paper we have
extended these results to higher powers in $\epsilon$.

Among the prospects for future developments and applications of
our results and methods are explicit three-loop checks of the
splitting and coefficient functions in semi-inclusive
$e^+e^-$-annihilation, for instance by computations of fixed-$N$
Mellin moments. Also QCD corrections to many other single scale
observables can be considered.

{\sc Form} files of our results can be obtained from the
preprint server {\tt http://arXiv.org} by downloading the source.
Furthermore they are available from the authors upon request.

%
%
{\bf{Acknowledgments:}}
We are grateful to A. Vogt for useful discussions.
A.M. acknowledges support by the Alexander von Humboldt Foundation.
The work of S.M. has been supported in part by the Helmholtz Gemeinschaft
under contract VH-NG-105.

Note added: The (un-)polarized coefficient functions up to two loops 
have recently been transformed to Mellin $N$-space in \cite{Blumlein:2006rr}.

\renewcommand{\theequation}{A.\arabic{equation}}
\setcounter{equation}{0}
%
%
\section*{Appendix A: Master Integrals}
\label{sec:appendix-A}
In this Appendix, we present the complete list of master
integrals, the corresponding difference equations in Mellin
$n$-space and the respective boundary conditions. We omit the
discussion of the so-called {\it purely virtual} contributions,
which are known since long. For the calculations of two-loop form
factors in QCD we refer to
Refs.~\cite{Matsuura:1988sm,Moch:2005id,Gehrmann:2005pd}.

Let us start with the master integrals ${\mbox{V}}_1(n), \dots,
{\mbox{V}}_5(n)$ for the so-called {\it real-virtual}
contributions. These masters can be defined through the following
object:
%
\begin{eqnarray}
\label{eq:rv-master}
{\lefteqn{
\left[i,\dots,j\right] = }}
\\
&& {e^{(3\gamma_{\rm e}\epsilon)}\over \pi^4}\,
\int d^dk d^dp_1 d^dp_2\,
\delta(p_1^2) \delta(p_2^2) \delta((q-p_1-p_2)^2)\,
\left(2q \cdot (q-p_1-p_2)\right)^n\, {1\over P_i\dots P_j}
\, ,
\nonumber
\end{eqnarray}
where $q \cdot q = 1$ and the propagators are
$P_1 = k^2,~
 P_2 =(q-p_1+k)^2,~
 P_3 = (q-k)^2,~
 P_4 = (p_1+p_2+k)^2,~
 P_5 = (p_1-k)^2,~
 P_6 = (p_2+k)^2,~
 P_7 = (q-p_2)^2$.

The $n$-dependent real-virtual masters are defined as:
${\mbox{V}}_1(n) = \left[1,2\right],~
 {\mbox{V}}_2(n) = \left[1,3\right],~
 {\mbox{V}}_3(n) = \left[1,4\right],~
 {\mbox{V}}_4(n) = \left[1,2,5\right],~
 {\mbox{V}}_5(n) = \left[1,2,5,6,7\right]$.
These masters satisfy the following difference equations:
\begin{eqnarray}
&&
         (n + 2 - 4 \* \epsilon) \* {\mbox{V}}_1(n)
       - (n + 1 - 3 \* \epsilon) \* {\mbox{V}}_1(n-1)
\, = \, 0 \: \: ,\label{eq:vm1-n}
\\[1ex]
&&
         (n + 2 - 3 \* \epsilon) \* {\mbox{V}}_2(n)
       - (n + 1 - 2 \* \epsilon) \* {\mbox{V}}_2(n-1)
\, = \, 0 \: \: ,\label{eq:vm2-n}
\\[1ex]
&&
         (n + 2 - 4 \* \epsilon) \* {\mbox{V}}_3(n)
       - (n + 1 - 2 \* \epsilon) \* {\mbox{V}}_3(n-1)
\, = \, 0 \: \: ,\label{eq:vm3-n}
\\[1ex]
&&
         (n + 1 - 2 \* \epsilon) \* {\mbox{V}}_4(n)
       - n \* {\mbox{V}}_4(n-1)
\, = \, \label{eq:vm4-n}
\\&& \mbox{}
       - (1 - 3 \* \epsilon) \* (1 - 2 \* \epsilon) \* {1 \over \epsilon} \* {\mbox{V}}_1(n-1)
       + (1 - 2 \* \epsilon)^2 \* {1 \over \epsilon} \* {\mbox{V}}_2(n-1)
\: \: ,
\nonumber\\[1ex]
&&
         (n - 1 - 4 \* \epsilon) \* {\mbox{V}}_5(n)
       - (n - 1 - 2 \* \epsilon) \* {\mbox{V}}_5(n-1)
\, = \, \label{eq:vm5-n}
\\&& \mbox{}
       - {{(1 - 3 \* \epsilon) \* (1 - 2 \* \epsilon) \* (n + 1 - 4 \* \epsilon)}
       \over {(n - 3 \* \epsilon) \* (n - 1 - 3 \* \epsilon)}}  \*
     (27 \* \epsilon^2  - 11 \* n \* \epsilon + 5 \* \epsilon - n + n^2)
       \* {1 \over \epsilon^2} \* {\mbox{V}}_1(n-1)
\nonumber\\&& \mbox{}
       + {{(1 - 2 \* \epsilon)^2 \* (n - 4 \* \epsilon) \* (n + 1 - 3 \* \epsilon)}
         \over {n - 2 \* \epsilon}} \* {1 \over \epsilon^2}
       \*{\mbox{V}}_2(n-1)
       + (n - 2 \* \epsilon) \* {\mbox{V}}_4(n-1)
\: \: .\nonumber
\end{eqnarray}
The boundary conditions ${\mbox{V}}_1(0), \dots, {\mbox{V}}_5(0)$
at $n=0$ read,
\begin{eqnarray}
  \lefteqn{  {\mbox{V}}_1(0) \,=\,  {\mbox{V}}_3(0) \,=\,}
\\&& \mbox{}
       {1 \over \epsilon} \* {1 \over 8}
      +{5 \over 4}
      +\epsilon \* \biggl(8 - {21 \over 16} \* \z2\biggr)
      +\epsilon^2 \* \biggl(42 - {23 \over 8} \* \z3 - {105 \over 8} \* \z2\biggr)
      +\epsilon^3 \* \biggl(198 - {115 \over 4} \* \z3 - 84 \* \z2 + {1017 \over 320} \* \z2^2\biggr)
\nonumber\\&& \mbox{}
      +\epsilon^4 \* \biggl(876 - {1053 \over 40} \* \z5 - 184 \* \z3 - 441 \* \z2
       + {483 \over 16} \* \z2 \* \z3
       + {1017 \over 32} \* \z2^2\biggr)
      +\epsilon^5 \* \biggl(3728 - {1053 \over 4} \* \z5
\nonumber\\&& \mbox{}
       - 966 \* \z3 + {529 \over 16} \* \z3^2
       - 2079 \* \z2 + {2415 \over 8} \* \z2 \* \z3 + {1017 \over 5} \* \z2^2 - {24737 \over 4480} \* \z2^3\biggr)
\: \: ,\label{eq:vm1}
\nonumber\\
  \lefteqn{  {\mbox{V}}_2(0) \,=\,}
\\&& \mbox{}
       {1 \over \epsilon} \* {1 \over 8}
      +{17 \over 16}
      +\epsilon \* \biggl({183 \over 32} - {17 \over 16} \* \z2\biggr)
      +\epsilon^2 \* \biggl({1597 \over 64} - {13 \over 8} \* \z3 - {289 \over 32} \* \z2\biggr)
      +\epsilon^3 \* \biggl({12359 \over 128} - {221 \over 16} \* \z3 - {3111 \over 64} \* \z2
\nonumber\\&& \mbox{}
       + {897 \over 320} \* \z2^2\biggr)
      +\epsilon^4 \* \biggl({88629 \over 256} - {303 \over 40} \* \z5 - {2379 \over 32} \* \z3 - {27149 \over 128} \* \z2
       + {221 \over 16} \* \z2 \* \z3 + {15249 \over 640} \* \z2^2\biggr)
\nonumber\\&& \mbox{}
      +\epsilon^5 \* \biggl({603871 \over 512} - {5151 \over 80} \* \z5 - {20761 \over 64} \* \z3
       + {169 \over 16} \* \z3^2
       - {210103 \over 256} \* \z2 + {3757 \over 32} \* \z2 \* \z3 + {164151 \over 1280} \* \z2^2
\nonumber\\&& \mbox{}
       - {12949 \over 4480} \* \z2^3\biggr)
\: \: ,\label{eq:vm2}
\nonumber\\
\lefteqn{  {\mbox{V}}_4(0) \,=\,}
\\&& \mbox{}
      - {1 \over \epsilon} \* {1 \over 4}
      - {11 \over 4} + {1 \over 2} \* \z2
      +\epsilon \* \biggl(-{77 \over 4} + {5 \over 2} \* \z3 + {49 \over 8} \* \z2\biggr)
      +\epsilon^2 \* \biggl(-{439 \over 4} + {93 \over 4} \* \z3 + {363 \over 8} \* \z2 - {3 \over 4} \* \z2^2\biggr)
\nonumber\\&& \mbox{}
      +\epsilon^3 \* \biggl(-{2229 \over 4} + {75 \over 2} \* \z5 + {583 \over 4} \* \z3 + {2141 \over 8} \* \z2
       - {131 \over 4} \* \z2 \* \z3 - {1857 \over 160} \* \z2^2\biggr)
      +\epsilon^4 \* \biggl(-{10527 \over 4} + {6303 \over 20} \* \z5
\nonumber\\&& \mbox{}
       + {3081 \over 4} \* \z3 - 45 \* \z3^2 + {11111 \over 8} \* \z2
       - {2317 \over 8} \* \z2 \* \z3 - {15147 \over 160} \* \z2^2 + {421 \over 80} \* \z2^3\biggr)
\: \: ,\label{eq:vm4}
\nonumber\\
\lefteqn{  {\mbox{V}}_5(0) \,=\, }
\\&& \mbox{}
       {1 \over \epsilon^4} \* {5 \over 8}
      +{1 \over \epsilon^3} \* {5 \over 4}
      +{1 \over \epsilon^2} \* \biggl({5 \over 2} - {133 \over 16} \* \z2\biggr)
      +{1 \over \epsilon} \* \biggl(5 - {133 \over 8} \* \z2 - {193 \over 8} \* \z3\biggr)
      +10 - {133 \over 4} \* \z2 + {5477 \over 320} \* \z2^2
\nonumber\\&& \mbox{}
       - {193 \over 4} \* \z3
      +\epsilon \* \biggl(20 + {4545 \over 16} \* \z2 \* \z3 - {133 \over 2} \* \z2 + {5477 \over 160} \* \z2^2
       - {193 \over 2} \* \z3 - {2303 \over 8} \* \z5\biggr)
      +\epsilon^2 \* \biggl(40
\nonumber\\&& \mbox{}
       + {4545 \over 8} \* \z2 \* \z3 - 133 \* \z2 + {5477 \over 80} \* \z2^2
       - {578731 \over 13440} \* \z2^3 - 193 \* \z3 + {5837 \over 16} \* \z3^2 - {2303 \over 4} \* \z5\biggr)
\: \: ,\label{eq:vm5} \nonumber
\end{eqnarray}
which we have given in terms of the Riemann
zeta-function consistently up to weight 6. Previously, the result
for ${\mbox{V}}_5(0)$ has been obtained to weight 4 in
Ref.~\cite{DeRidder:2003bm}.

Next we present the master integrals ${\mbox{R}}_1(n), \dots,
{\mbox{R}}_6(n)$ from the so-called {\it real-real} contributions.
These masters can be defined through the following object:
%
\begin{eqnarray}
\label{eq:rr-master}
\left\{i,\dots,j\right\}  &=&
{e^{(3\gamma_{\rm e}\epsilon)}\over \pi^3}\,
\int d^dp_1 d^dp_2 d^dp_3\,
\delta(p_1^2) \delta(p_2^2) \delta(p_3^2) \delta((q-p_1-p_2-p_3)^2)
\\
&& \times \left(2q \cdot (q-p_1-p_2-p_3)\right)^n\, {1\over Q_i\dots Q_j}
\, ,
\nonumber
\end{eqnarray}
where $q \cdot q = 1$ and the propagators are
$Q_1 = (q-p_3)^2,~
 Q_2 = (q-p_2)^2,~
 Q_3 = (q-p_1-p_3)^2,~
 Q_4 = (q-p_2-p_3)^2,~
 Q_5 = (q-p_1-p_2)^2,~
 Q_6 = (p_1+p_2)^2,~
 Q_7 = (p_1+p_3)^2,~
 Q_8 = (p_2+p_3)^2$.

The $n$-dependent real-real masters are defined as:
${\mbox{R}}_1(n) = \left\{ - \right\},~
 {\mbox{R}}_2(n) = \left\{1,2\right\},~
 {\mbox{R}}_3(n) = \left\{2,6\right\},~
 {\mbox{R}}_4(n) = \left\{1,2,3,5\right\},~
 {\mbox{R}}_5(n) = \left\{1,2,6,7\right\},~
 {\mbox{R}}_6(n) = \left\{3,4,7,8\right\}$
and satisfy the following difference equations in $n$
together with boundary conditions at $n=0$:
\begin{eqnarray}
&&
         (n + 3 - 4 \* \epsilon) \* {\mbox{R}}_1(n)
       - (n  + 1 - 2 \* \epsilon) \* {\mbox{R}}_1(n-1)
\, = \, 0 \: \: ,\label{eq:rm1-n}
\\[1ex]
&&
         (n + 1 - 2 \* \epsilon) \* (n + 2 - 6 \* \epsilon)  \*  {\mbox{R}}_2(n)
       - (n-1) \* (n - 4 \* \epsilon)  \*  {\mbox{R}}_2(n-2)
\, = \, \label{eq:rm2-n}
\\&& \mbox{}
         2 \* {{(1-\epsilon) \* (1 - 3 \* \epsilon) \* (2 - 3 \* \epsilon) \*
           (2 \* n + 1 - 6 \* \epsilon)} \over {(n - 3 \* \epsilon) \* (n + 1 - 3 \* \epsilon)}}\,
       \* {\mbox{R}}_1(n-2)
\: \: ,
\nonumber\\[1ex]
&&
         (n + 1 - 2 \* \epsilon) \* {\mbox{R}}_3(n)
       - n \* {\mbox{R}}_3(n-1)
\, = \,
         - {{(1-3 \* \epsilon) \* (2-3 \* \epsilon) \* (n+2-4 \* \epsilon) }
           \over {n + 1 - 3 \* \epsilon}}\, \* {1 \over \epsilon}\, \*  {\mbox{R}}_1(n-1)
\: \: , \label{eq:rm3-n}
\\[1ex]
&&
         (n - 1 - 4 \* \epsilon)  \*   {\mbox{R}}_4(n)
       + (n - 1 - 2 \* \epsilon)  \*   {\mbox{R}}_4(n-1)
\, = \, \label{eq:rm4-n}
\\&& \mbox{}
       4 \*  {{(1-3 \* \epsilon) \* (2-3 \* \epsilon) \* (n+2-4 \* \epsilon)} \over
         {(n - 2 \* \epsilon) \* (n - 3 \* \epsilon) \* (n - 4 \* \epsilon) \*
           (n - 1 - 3 \* \epsilon) \* (n - 1 - 4 \* \epsilon) \* (n + 1 - 3 \* \epsilon)}} \*
       \biggl(3240 \* \epsilon^6-5058 \* n \* \epsilon^5
\nonumber\\&& \mbox{}
        -1098 \* \epsilon^5+3282 \* n^2 \* \epsilon^4-279 \* \epsilon^4+1407 \* n \* \epsilon^4
        +52 \* \epsilon^3+269 \* n \* \epsilon^3-1135 \* n^3 \* \epsilon^3-707 \* n^2 \* \epsilon^3
\nonumber\\&& \mbox{}
        -41 \* n \* \epsilon^2+174 \* n^3 \* \epsilon^2
        +5 \* \epsilon^2+221 \* n^4 \* \epsilon^2-98 \* n^2 \* \epsilon^2+16 \* n^3 \* \epsilon
        -21 \* n^4 \* \epsilon+11 \* n^2 \* \epsilon-n \* \epsilon-23 \* n^5 \* \epsilon
\nonumber\\&& \mbox{}
        -n^4-n^3+n^5+n^6 \biggr)
       \* {1 \over \epsilon^2}\, \* {\mbox{R}}_1(n-1)
       + {{( n + 1 - 2 \* \epsilon) \* ( n + 2 - 6 \* \epsilon) \* ( n - 6 \* \epsilon)} \over {n - 4 \* \epsilon}}
       \* {1 \over \epsilon}\, \*  {\mbox{R}}_2(n)
\nonumber\\&& \mbox{}
       - {{(n - 2 \* \epsilon) \* ( n + 1 - 6 \* \epsilon) \* ( n - 1 - 6 \* \epsilon) } \over {n - 1 - 4 \* \epsilon}}
       \* {1 \over \epsilon}\, \*  {\mbox{R}}_2(n-1)
\: \: ,
\nonumber\\[1ex]
&&
         (n - 1 - 4 \* \epsilon) \* {\mbox{R}}_5(n)
       - (n - 1 - 2 \* \epsilon) \* {\mbox{R}}_5(n-1)
\, = \, \label{eq:rm5-n}
\\&& \mbox{}
        2 \* {{(1-3 \* \epsilon) \* (2-3 \* \epsilon) \* (n+2-4 \* \epsilon)} \over
         {(n - 3 \* \epsilon) \* (n - 4 \* \epsilon) \* (n - 1 - 4 \* \epsilon) \* (n + 1 - 3 \* \epsilon)}} \ \*
       \biggl(120 \* \epsilon^4-154 \* n \* \epsilon^3-142 \* \epsilon^3+71 \* n^2 \*
\epsilon^2 
\nonumber\\&& \mbox{}
        +104 \* n \* \epsilon^2
        +23 \* \epsilon^2-10 \* n \* \epsilon-25 \* n^2 \* \epsilon-14 \* n^3 \* \epsilon
        -\epsilon+n^2+n^4+2 \* n^3\biggr)
        \* {1 \over \epsilon^2}\, \* {\mbox{R}}_1(n-1)
\nonumber\\&& \mbox{}
       + 2 \* (n-2 \* \epsilon) \* {\mbox{R}}_3(n-1)
       + {{( n + 1 - 2 \* \epsilon) \* ( n + 2 - 6 \* \epsilon) \* ( n - 6 \* \epsilon)} \over {n - 4 \* \epsilon}}
       \* {1 \over \epsilon}\, \*  {\mbox{R}}_2(n)
\nonumber\\&& \mbox{}
       + {{(n - 2 \* \epsilon) \* ( n + 1 - 6 \* \epsilon) \* ( n - 1 - 6 \* \epsilon) } \over {n - 1 - 4 \* \epsilon}}
       \* {1 \over \epsilon}\, \*  {\mbox{R}}_2(n-1)
\: \: ,
\nonumber\\[1ex]
&&
         (n - 1 - 4 \* \epsilon) \* {\mbox{R}}_6(n)
       - (n - 1 - 2 \* \epsilon) \* {\mbox{R}}_6(n-1)
\, = \, 0 \: \: .\label{eq:rm6-n} 
\end{eqnarray}
\begin{eqnarray}
  \lefteqn{ {\mbox{R}}_1(0) \,=\,}
\\&& \mbox{}
       {1 \over 96}
      +\epsilon \* {71 \over 576}
      +\epsilon^2 \* \biggl({3115 \over 3456} - {7 \over 64} \* \z2\biggr)
      +\epsilon^3 \* \biggl({109403 \over 20736} - {29 \over 96} \* \z3
       - {497 \over 384} \* \z2\biggr)
      +\epsilon^4 \* \biggl({3386467 \over 124416} - {2059 \over 576} \* \z3
\nonumber\\&& \mbox{}
       - {21805 \over 2304} \* \z2
       + {291 \over 1280} \* \z2^2\biggr)
      +\epsilon^5 \* \biggl({96885467 \over 746496} - {421 \over 160} \* \z5 - {90335 \over 3456} \* \z3
        - {765821 \over 13824} \* \z2 + {203 \over 64} \* \z2 \* \z3
\nonumber\\&& \mbox{}
        + {6887 \over 2560} \* \z2^2\biggr)
      +\epsilon^6 \* \biggl({2631913075 \over 4478976} - {29891 \over 960} \* \z5
        - {3172687 \over 20736} \* \z3 + {841 \over 192} \* \z3^2
        - {23705269 \over 82944} \* \z2
\nonumber\\&& \mbox{}
        + {14413 \over 384} \* \z2 \* \z3 + {60431 \over 3072} \* \z2^2
        - {15089 \over 53760} \* \z2^3\biggr)
\: \: ,\label{eq:rm1}
\nonumber\\
  \lefteqn{ {\mbox{R}}_2(0) \,=\,}
\\&& \mbox{}
      - {1 \over 8} + {1 \over 8} \* \z2
      +\epsilon \* \biggl(- {7 \over 4} + {9 \over 8} \* \z3 + {7 \over 8} \* \z2\biggr)
      +\epsilon^2 \* \biggl(- {119 \over 8} + {63 \over 8} \* \z3 + {87 \over 16} \* \z2
       + {97 \over 80} \* \z2^2\biggr)
      +\epsilon^3 \* \biggl(- {199 \over 2}
\nonumber\\&& \mbox{}
       + {207 \over 8} \* \z5 + {163 \over 4} \* \z3
       + {139 \over 4} \* \z2 - {211 \over 16} \* \z2 \* \z3 + {679 \over 80} \* \z2^2 \biggr)
      +\epsilon^4 \* \biggl(- {4617 \over 8} + {1449 \over 8} \* \z5 + {1585 \over 8} \* \z3
\nonumber\\&& \mbox{}
         - {45 \over 2} \* \z3^2 + {3445 \over 16} \* \z2 - {1477 \over 16} \* \z2 \* \z3
         + {11931 \over 320} \* \z2^2 + {1141 \over 320} \* \z2^3\biggr)
\: \: ,\label{eq:rm2}
\nonumber\\
  \lefteqn{ {\mbox{R}}_3(0) \,=\,}
\\&& \mbox{}
      -{1 \over \epsilon} \* {1 \over 8}
      -{11 \over 8}
      +\epsilon \* \biggl(- {77 \over 8} + {21 \over 16} \* \z2\biggr)
      +\epsilon^2 \* \biggl(- {439 \over 8} + {29 \over 8} \* \z3 + {231 \over 16} \* \z2\biggr)
      +\epsilon^3 \* \biggl(- {2229 \over 8} + {319 \over 8} \* \z3
\nonumber\\&& \mbox{}
       + {1617 \over 16} \* \z2
       - {873 \over 320} \* \z2^2\biggr)
      +\epsilon^4 \* \biggl(-{10527 \over 8} + {1263 \over 40} \* \z5 + {2233 \over 8} \* \z3
       + {9219 \over 16} \* \z2 - {609 \over 16} \* \z2 \* \z3
\nonumber\\&& \mbox{}
       - {9603 \over 320} \* \z2^2\biggr)
      +\epsilon^5 \* \biggl( - {47389 \over 8} + {13893 \over 40} \* \z5
       + {12731 \over 8} \* \z3 - {841 \over 16} \* \z3^2 + {46809 \over 16} \* \z2
       - {6699 \over 16} \* \z2 \* \z3
\nonumber\\&& \mbox{}
       - {67221 \over 320} \* \z2^2 + {15089 \over 4480} \* \z2^3\biggr)
\: \: ,\label{eq:rm3}
\nonumber\\[6ex]
\lefteqn{ {\mbox{R}}_4(0) \,=\, {\mbox{R}}_5(0) \,=\,}
\\&& \mbox{}
       {1 \over \epsilon^4} \* {3 \over 32}
      +{1 \over \epsilon^3} \* {3 \over 16}
      +{1 \over \epsilon^2} \* \biggl({3 \over 8} - {83 \over 64} \* \z2\biggr)
      +{1 \over \epsilon} \* \biggl({3 \over 4} - {185 \over 32} \* \z3
       - {83 \over 32} \* \z2\biggr)
      + {3 \over 2} - {185 \over 16} \* \z3 - {83 \over 16} \* \z2 - {457 \over 256} \* \z2^2
\nonumber\\&& \mbox{}
      +\epsilon \* \biggl(3-{15899 \over 160} \* \z5 - {185 \over 8} \* \z3 - {83 \over 8} \* \z2
       + {4137 \over 64} \* \z2 \* \z3 - {457 \over 128} \* \z2^2\biggr)
      +\epsilon^2 \* \biggl(6-{15899 \over 80} \* \z5 - {185 \over 4} \* \z3
\nonumber\\&& \mbox{}
       + {6667 \over 64} \* \z3^2
       - {83 \over 4} \* \z2 + {4137 \over 32} \* \z2 \* \z3
       - {457 \over 64} \* \z2^2 - {692333 \over 53760} \* \z2^3\biggr)
\: \: ,\label{eq:rm5}
\nonumber\\
  \lefteqn{ {\mbox{R}}_6(0) \,=\,}
\\&& \mbox{}
       {1 \over \epsilon^4} \* {5 \over 8}
      +{1 \over \epsilon^3} \* {5 \over 4}
      +{1 \over \epsilon^2} \* \biggl({5 \over 2} - {105 \over 16} \* \z2\biggr)
      +{1 \over \epsilon} \* \biggl(5 - {141 \over 8} \* \z3
       - {105 \over 8} \* \z2\biggr)
      +10 - {141 \over 4} \* \z3 - {105 \over 4} \* \z2
\nonumber\\&& \mbox{}
      + {969 \over 64} \* \z2^2
      +\epsilon \* \biggl(20 - {1119 \over 8} \* \z5
       - {141 \over 2} \* \z3
       - {105 \over 2} \* \z2 + {2961 \over 16} \* \z2 \* \z3 + {969 \over 32} \* \z2^2\biggr)
      +\epsilon^2 \* \biggl(40 - {1119 \over 4} \* \z5
\nonumber\\&& \mbox{}
       - 141 \* \z3 + {4005 \over 16} \* \z3^2
       - 105 \* \z2 + {2961 \over 8} \* \z2 \* \z3 + {969 \over 16} \* \z2^2
       - {15121 \over 896} \* \z2^3\biggr)
\: \: .\label{eq:rm6} \nonumber 
\end{eqnarray}
Again we have given the results consistently up to weight 6 in the
Riemann zeta-function. The difference equation for
${\mbox{R}}_2(n)$ which is of second order needs actually two
boundary conditions. The value for ${\mbox{R}}_2(1)$ is obtained
from an additional fixed-$n$ reduction
\begin{eqnarray}
  \label{eq:rm2-1}
  {\mbox{R}}_2(1) \,=\, {{ 3 - 4 \* \epsilon}\over {1 - 2 \* \epsilon}}\, {\mbox{R}}_1(0)
\: \: .
\end{eqnarray}
The values for ${\mbox{R}}_4(0)$, ${\mbox{R}}_5(0)$ and
${\mbox{R}}_6(0)$ have been obtained before up to weight 4 in
Ref.~\cite{DeRidder:2003bm}.

\renewcommand{\theequation}{B.\arabic{equation}}
\setcounter{equation}{0}
%
%
\section*{Appendix B: The exact Mellin-space results}
\label{sec:appendix-B}
Here we give the exact Mellin-$N$ expressions for the
time-like splitting functions and the coefficient functions
$c_{T}$, $c_{L}$ and $c_{A}$ up to second order in $\ar = \as /
(4\pi)$, expressed in terms of harmonic sums, recursively
defined by \cite{Vermaseren:1998uu}
\begin{equation}
  \label{eq:Ssum}
  S_{\pm m_1,m_2,\ldots,m_k}(N) \:\: = \:\: \sum_{i=1}^{N}\:
  \frac{(\pm 1)^{i}}{i^{\, m_1}}\: S_{m_2,\ldots,m_k}(i)
  \:\: , \quad\quad  S(N) \:\:  = \:\: 1 \:\: ,
\end{equation}
and we employ the notation
\begin{equation}
  \Npm \, S_{\vec{m}} \: = \: S_{\vec{m}}(N \pm 1) \:\: , \quad\quad
  \Npmi\, S_{\vec{m}} \: = \: S_{\vec{m}}(N \pm i) \:\: .
\end{equation}

The well known results for the LO anomalous dimensions (cf. Eq.~(\ref{eq:defP}))
in $N$-space are:
\begin{eqnarray}
&& \gamma^{(0)}_{\rm qq}(N) \, = \,
 \colour4colour{\cf}\*\Bigg\{-3+(2\*\Nminus+2\*\Nplus)\*\S(1)\Bigg\}
\: \: ,
\\[2ex]
&& \gamma^{(0)}_{\rm qg}(N) \, = \,
 \colour4colour{\nf}\*\Bigg\{2\*\Nminus+8\*\Nplus-4\*\Nplustwo-6\Bigg\}\*\S(1)
\: \: ,
\\[2ex]
&& \gamma^{(0)}_{\rm gq}(N) \, = \,
 \colour4colour{\cf}\*\Bigg\{-8\*\Nminus+4\*\Nminustwo-2\*\Nplus+6\Bigg\}\*\S(1)
\: \: ,
\\[2ex]
&& \gamma^{(0)}_{\rm gg}(N) \, = \,
 \colour4colour{\ca}\*\Bigg\{
 -{11\over3}
 +(-8\*\Nminus+4\*\Nminustwo-8\*\Nplus+4\*\Nplustwo+12)\*\S(1)
 \Bigg\}
+ {2\over 3}\*\colour4colour{\nf}
\: \: .
\end{eqnarray}

For the NLO anomalous dimensions (cf. Eq.~(\ref{eq:defP})) a
distinction between even and odd Mellin moments is in order. We
give the respective expressions for the even Mellin moments of
$\gamma^{(1)+}_{\rm ns}$ in Eq.~(\ref{eq:gTqq1}) and $\gamma^{(1)
\rm s}_{\rm qq}$, $\gamma^{(1)}_{\rm qg}$, $\gamma^{(1)}_{\rm
gq}$, $\gamma^{(1)}_{\rm gg}$ in
Eqs.~(\ref{eq:gTqqps1})--(\ref{eq:gTgg1}) and for the odd Mellin
moments of $\gamma^{(1)-}_{\rm ns}$ in Eq.~(\ref{eq:gTqq1m}). Note
also, that we do not obtain $\gamma_{\,\rm qg}^{\,(1)}$ and
$\gamma_{\,\rm gg}^{\,(1)}$ since we only consider vector boson
decays (see discussion in Section~\ref{sec:results}).
Nevertheless, we also quote these quantities here for
completeness.
\begin{eqnarray}
\label{eq:gTqq1}
\lefteqn{ \gamma^{(1)+}_{\rm ns}(N) \, = \, }
\\&&
\colour4colour{\cf^2}\* \Bigg\{
\Bigg(8-
(16\*\z2+4)\*\Nplus-
(16\*\z2+4)\*\Nminus\Bigg)\*\S(1) -
\left(4\*\Nplus+8\right)\*\S(2)-
(4\*\Nminus+28\*\Nplus)\*\S(3)+
\nonumber\\&&
(-16\*\Nminus-16\*\Nplus)\*\Ss(1,-2)+
(8\*\Nminus+8\*\Nplus)\*\Ss(1,2)+
(8\*\Nminus+8\*\Nplus)\*\Ss(2,1)+
16\*\S(-3) +
24\*\z2-{3\over 2}
\Bigg\} +
\nonumber\\&&
\colour4colour{\cf\*\ca}\*
\Bigg\{
-8\*\S(-3)
+\Bigg({302\over 9}\*\Nplus-{112\over 3}
+{302\over 9}\*\Nminus\Bigg)\*\S(1)
+\Bigg(-{22\over 3}\*\Nminus
-{22\over 3}\*\Nplus\Bigg)\*\S(2)
+8\*\Nplus\*\S(3) -
\nonumber\\&&
{17\over 6}
+ (8\*\Nminus+8\*\Nplus)\*\Ss(1,-2)
\Bigg\} +
\nonumber\\&&
\colour4colour{\cf\*\nf}\*\Bigg\{\Bigg(
-{44\over 9}\*\Nplus
-{44\over 9}\*\Nminus
+{16\over 3}\Bigg)\*\S(1)
+{1\over 3}
+\Bigg({4\over 3}\*\Nminus
+{4\over 3}\*\Nplus\Bigg)\*\S(2)
\Bigg\}
\: \: ,\nonumber\\[2ex]
\label{eq:gTqq1m}
\lefteqn{ \gamma^{(1)-}_{\rm ns}(N) \, = \,}
\\&&
\colour4colour{\cf^2}\*\Bigg\{
16\*\S(-3)
-((16\*\z2+36)\*\Nplus
-72
+(16\*\z2+36)\*\Nminus)\*\S(1)
+(16\*\Nminus
-20\*\Nplus
-8)\*\S(2) -
\nonumber\\&&
(20\*\Nminus+12\*\Nplus)\*\S(3)
+(8\*\Nplus+8\*\Nminus)\*\Ss(2,1)
-(16\*\Nminus+16\*\Nplus)\*\Ss(1,-2)
+(8\*\Nplus+8\*\Nminus)\*\Ss(1,2) -
\nonumber\\&&
{3\over 2} + 24\*\z2\Bigg\} +
\nonumber\\&&
\colour4colour{\cf\*\ca}\*\Bigg\{
-8\*\S(-3)
+\Bigg({446\over 9}\*\Nminus
-{208\over 3}
+{446\over 9}\*\Nplus\Bigg)\*\S(1)
+\Bigg(-{46\over 3}\*\Nminus
+{2\over 3}\*\Nplus\Bigg)\*\S(2)
+8\*\Nminus\*\S(3) -
\nonumber\\&&
{17\over 6}
+(8\*\Nplus+8\*\Nminus)\*\Ss(1,-2)
\Bigg\} +
\nonumber\\&&
\colour4colour{\cf\*\nf}\*\Bigg\{\Bigg(
-{44\over 9}\*\Nplus
-{44\over 9}\*\Nminus
+{16\over 3}\Bigg)\*\S(1)
+{1\over 3}
+\Bigg({4\over 3}\*\Nplus
+{4\over 3}\*\Nminus\Bigg)\*\S(2)\Bigg\}
\: \: ,\nonumber\\[2ex]
\label{eq:gTqqps1}
\lefteqn{ \gamma^{(1) \rm s}_{\rm qq}(N) \, = \, }
\\&&
\colour4colour{\cf\*\nf}\*
\Bigg\{\Bigg(
-{208\over 9}\*\Nminus
-{224\over 9}\*\Nplustwo
+48
-{80\over 9}\*\Nminustwo
+{80\over 9}\*\Nplus\Bigg)\*\S(1) +
\nonumber\\&&
\Bigg(
-{76\over 3}\*\Nplus
-{32\over 3}\*\Nplustwo
+16
+20\*\Nminus\Bigg)\*\S(2)
+(8\*\Nminus
-8\*\Nplus)\*\S(3)\Bigg\}
\: \: ,
\nonumber\\[2ex]
\label{eq:gTqg1}
\lefteqn{ \gamma^{(1)}_{\rm qg}(N) \, = \, }
\\&&
\colour4colour{\cf\*\nf}\*\Bigg\{(
-86\*\Nplus
+40\*\Nplustwo
+70
-24\*\Nminus)\*\S(1)
+(10\*\Nminus+8\*\Nplustwo-18)\*\S(2)
+ \Bigg(16\*\Nplustwo
+12 -
\nonumber\\&&
4\*\Nminus
-24\*\Nplus\Bigg)\*\S(3)
+(16\*\Nplus-20-8\*\Nplustwo+12\*\Nminus)\*\Ss(1,1)
+\Bigg(96\*\Nplus+24\*\Nminus-48\*\Nplustwo-
\nonumber\\&&
72\Bigg)\*\Ss(1,2)
+(24-8\*\Nminus+16\*\Nplustwo-32\*\Nplus)\*\Ss(2,1)
+(24-8\*\Nminus+16\*\Nplustwo-32\*\Nplus)\*\Sss(1,1,1)
\Bigg\} +
\nonumber\\&&
\colour4colour{\ca\*\nf}\*\Bigg\{\Bigg(
\Bigg({44\over 3}-16\*\z2\Bigg)\*\Nminus -
\Bigg(64\*\z2-{364\over 3}\Bigg)\*\Nplus+
\Bigg(32\*\z2-{712\over 9}\Bigg)\*\Nplustwo
+48\*\z2 -
\nonumber\\&&
{80\over 9}\*\Nminustwo
- 48\Bigg)\*\S(1)
+ \Bigg(-{16\over 3}\*\Nplustwo
+40
-40\*\Nplus
+{16\over 3}\*\Nminus\Bigg)\*\S(2)
+(8\*\Nminus-48\*\Nplus+40)\*\S(3) +
\nonumber\\&&
(24-8\*\Nminus +
16\*\Nplustwo
-32\*\Nplus)\*\Ss(1,-2)
+\Bigg(-{80\over 3}\*\Nplus
+{40\over 3}\*\Nplustwo
+28
-{44\over 3}\*\Nminus\Bigg)\*\Ss(1,1) +
\nonumber\\&&
(48-16\*\Nminus-64\*\Nplus
+32\*\Nplustwo )\*\Ss(1,2)
+(64\*\Nplus+16\*\Nminus-48-32\*\Nplustwo)\*\Ss(2,1) +
\nonumber\\&&
(8\*\Nminus+32\*\Nplus-16\*\Nplustwo-24)\*\Sss(1,1,1)
\Bigg\} +
\nonumber\\&&
\colour4colour{\nf^2}\*\Bigg\{
\Bigg(8-{40\over 9}\*\Nminus
+{32\over 9}\*\Nplustwo
-{64\over9}\*\Nplus\Bigg)\*\S(1)
+\Bigg(-8+{8\over 3}\*\Nminus
-{16\over 3}\*\Nplustwo
+{32\over 3}\*\Nplus\Bigg)\*\S(2) +
\nonumber\\&&
\Bigg(-8+{8\over 3}\*\Nminus
-{16\over 3}\*\Nplustwo
+{32\over 3}\*\Nplus\Bigg)\*\Ss(1,1)\Bigg\}
\: \: ,\nonumber\\[2ex]
\label{eq:gTgq1}
\lefteqn{ \gamma^{(1)}_{\rm gq}(N) \, = \, }
\\&&
\colour4colour{\cf^2}\*\Bigg\{
\Bigg(( 16\*\z2-18)\*\Nplus
+( 64\*\z2-2)\*\Nminus
-48\*\z2+20-32\*\z2\*\Nminustwo\Bigg)\*\S(1)
+\Bigg( 32\*\Nminus
-34 +
\nonumber\\&&
2\*\Nplus\Bigg)\*\S(2)
+(8\*\Nminus-12+4\*\Nplus)\*\S(3)
+(8\*\Nplus-8)\*\Ss(1,1)
+\Bigg(16\*\Nplus-32\*\Nminustwo
+64\*\Nminus -
\nonumber\\&&
48\Bigg)\*\Ss(1,2)
+(48-64\*\Nminus+32\*\Nminustwo-16\*\Nplus)\*\Ss(2,1) +
(24-8\*\Nplus+16\*\Nminustwo-32\*\Nminus)\*\Sss(1,1,1)\Bigg\} +
\nonumber\\&&
\colour4colour{\cf\*\ca}\*\Bigg\{
\Bigg({68\over 9}\*\Nminustwo
+{176\over 9}\*\Nplustwo
+{112\over 9}\*\Nminus-24
-{140\over 9}\*\Nplus\Bigg)\*\S(1)
+\Bigg({32\over 3}\*\Nplustwo
+{76\over 3}\*\Nplus+24\*\Nminustwo -
\nonumber\\&&
4 - 56\*\Nminus\Bigg)\*\S(2)
+(24\*\Nplus+16\*\Nminus-8-32\*\Nminustwo)\*\S(3)
+(32\*\Nminus-16\*\Nminustwo+8\*\Nplus-24)\*\Ss(1,-2) +
\nonumber\\&&
(8-8\*\Nplus)\*\Ss(1,1)
+(48\*\Nminustwo-24\*\Nplus+72-96\*\Nminus)\*\Ss(1,2) +
(32\*\Nminus-16\*\Nminustwo+8\*\Nplus-24)\*\Ss(2,1) +
\nonumber\\&&
(-24+32\*\Nminus-16\*\Nminustwo+8\*\Nplus)\*\Sss(1,1,1)\Bigg\}
\: \: ,\nonumber\\[2ex]
\label{eq:gTgg1}
\lefteqn{ \gamma^{(1)}_{\rm gg}(N)  \, = \,}
\\&&
\colour4colour{\cf\*\nf}\*\Bigg\{
\Bigg(32+{328\over 9}\*\Nplustwo
-{256\over 9}\*\Nminus
+{184\over 9}\*\Nminustwo
-{544\over 9}\*\Nplus\Bigg)\*\S(1) +
\nonumber\\&&
\Bigg({32\over 3}\*\Nplustwo
-8
-{28\over 3}\*\Nminus
-{32\over 3}\*\Nminustwo
+{52\over 3}\*\Nplus\Bigg)\*\S(2)
+2
+(8\*\Nminus-8\*\Nplus)\*\S(3)\Bigg\} +
\nonumber\\&&
\colour4colour{\ca^2}\*\Bigg\{
8\*\S(-3)+
\Bigg(\Bigg({218\over 9}+64\*\z2\Bigg)\*\Nplus
-32\*\z2\*\Nplustwo+
\Bigg({218\over 9}
+64\*\z2\Bigg)\*\Nminus
-96\*\z2
-{56\over 3} -
\nonumber\\&&
32\*\z2\*\Nminustwo\Bigg)\*\S(1)
+\Bigg(-{88\over 3}\*\Nplustwo
-40
+{88\over 3}\*\Nminustwo
+{76\over 3}\*\Nplus
-{44\over 3}\*\Nminus\Bigg)\*\S(2) +
\Bigg(-16\*\Nplustwo -
\nonumber\\&&
32\*\Nminustwo - 72
+16\*\Nminus
+80\*\Nplus \Bigg)\*\S(3)
+(32\*\Nplus
+32\*\Nminus-16\*\Nplustwo
-16\*\Nminustwo
-48)\*\Ss(1,-2) +
\nonumber\\&&
(48-32\*\Nplus +
16\*\Nplustwo-32\*\Nminus+16\*\Nminustwo )\*\Ss(1,2) +
\nonumber\\&&
(48-32\*\Nplus+16\*\Nplustwo-32\*\Nminus+16\*\Nminustwo)\*\Ss(2,1)
+{88\over 3}\*\z2
-{32\over 3}\Bigg\} +
\nonumber\\&&
\colour4colour{\ca\*\nf}\*\Bigg\{
\Bigg({56\over 3}\*\Nminus
-{92\over 9}\*\Nplustwo
-{64\over 3}
-{92\over 9}\*\Nminustwo
+{56\over 3}\*\Nplus\Bigg)\*\S(1) +
\nonumber\\&&
\Bigg(-8\*\Nminus
+{16\over 3}\*\Nminustwo
+{16\over 3}\*\Nplustwo
-{40\over 3}\*\Nplus
+16\Bigg)\*\S(2)
+{8\over 3}
-{16\over 3}\*\z2 \Bigg\}
\: \: .\nonumber
\end{eqnarray}

The ${\cal O}(\as)$  coefficient functions in $N$-space read:
\begin{eqnarray}
&& c_{T,\rm q}^{(1)}(N) \, = \,
\colour4colour{\cf}\*\Bigg\{
(-3\*\Nplus+6)\*\S(1)
+(4\*\Nminus+4\*\Nplus)\*\S(2)
-9
+(2\*\Nminus
+2\*\Nplus)\*\Ss(1,1)
\Bigg\}
\: \: ,
\\[2ex]
&& c_{L,\rm q}^{(1)}(N) \, = \,
\colour4colour{\cf}\*(2-2\*\Nminus)\*\S(1)
\: \: ,
\\[2ex]
&& c_{A,\rm q}^{(1)}(N) - c_{T,\rm q}^{(1)}(N) \, = \,
\colour4colour{\cf}\*(2\*\Nminus+2\*\Nplus-4)\*\S(1)
\: \: ,
\\[2ex]
&& c_{T,\rm g}^{(1)}(N) \, = \,
\colour4colour{\cf}\*\Bigg\{
(-16\*\Nminus+8\*\Nminustwo+8)\*\S(1)
+(16\*\Nminustwo-8\*\Nplus+24-32\*\Nminus)\*\S(2)
\\&&
+(8\*\Nminustwo-4\*\Nplus+12-16\*\Nminus)\*\Ss(1,1)
\Bigg\}
\: \: ,\nonumber
\\[2ex]
&& c_{L,\rm g}^{(1)}(N) \, = \,
\colour4colour{\cf}\*(16\*\Nminus-8\*\Nminustwo-8)\*\S(1)
\: \: .
\end{eqnarray}

Finally, for the ${\cal O}(\as^2)$ coefficient functions we give
the even Mellin moments of $c_{I,\rm ns}^{(2)}$, $c_{I,\rm
ps}^{(2)}$, $c_{I,\rm g}^{(2)}$ with $I=T,L$ in
Eqs.~(\ref{eq:cTeq2})--(\ref{eq:cLeg2}) and the odd Mellin moments
of $c_{A,\rm ns}^{(2)}$ in Eq.~(\ref{eq:cAeq2}).
\begin{eqnarray}
\label{eq:cTeq2}
\lefteqn{ c_{T,\rm ns}^{(2)}(N) \, = \,}
\\&&
\delta(N-2)\*\Bigg[
\colour4colour{\cf^2}\*\Bigg\{-{560\over 9}\*\z2-{704\over 15}\*\z3+{145517\over 810}\Bigg\} +
\nonumber\\&&
\colour4colour{\cf\*\ca}\*\Bigg\{-{52\over 9}\*\z2-{316\over 5}\*\z3+{127349\over 810}\Bigg\} +
\colour4colour{\cf\*\nf}\*\Bigg\{-{2354\over 81}+8\*\z3+{16\over 9}\*\z2\Bigg\}
\Bigg] +
\nonumber\\&&
\theta(N-3)\*\Bigg[
\colour4colour{\cf^2}\*\Bigg\{59\*\z2-12\*\z3+72\*\S(-4)-16\*
\S(-2)+16\*\Ss(-3,1)+48\*\Ss(-2,-2)+{331\over 8} + \Bigg(\Bigg(-{118\over 5}+
\nonumber\\&&
12\*\z2\Bigg)\*\Nplus
-{48\over 5}\*\Nplustwo
+\Bigg(4\*\z2-{282\over 5} \Bigg)\*\Nminus-26
+{48\over 5}\*\Nminustwo\Bigg)\*\S(2)
+\Bigg(\Bigg( 8\*\z3 -{75\over 2} - 12\*\z2\Bigg)\*\Nplus+
\nonumber\\&&
{48\over 5}\*\Nplustwo+(-63-12\*\z2+8\*\z3)\*\Nminus-48\*\z3
+{48\over 5}\*\Nminustwo+{279\over 5}\Bigg)\*\S(1)
+ \Bigg({48\over 5}\*\Nplusthree-36 - {48\over 5}\*\Nplustwo-
\nonumber\\&&
30\*\Nplus\Bigg)\*\S(3)-(22\*\Nminus+138\*\Nplus)\*\S(4)
-(16\*\Nminus+16\*\Nplus)\*\Sss(1,-2,1)
+ (32-48\*\Nminus-48\*\Nplus)\*\Sss(1,1,-2)+
\nonumber\\&&
(-16-16\*\Nminus-16\*\Nplus)\*\Ss(1,-3)
+ \Bigg(-8\*\Nminus-{48\over 5}\*\Nminustwo +16
- {48\over 5}\*\Nplustwo+{48\over 5}\*\Nplusthree
+ {48\over 5}\*\Nminusthree-
\nonumber\\&&
8\*\Nplus\Bigg)\*\Ss(1,-2)+(56-28\*\Nplus-4\*\Nminus)\*\Ss(1,2)
+\Bigg( 64 - (32\*\z2+51)\*\Nplus -(32\*\z2 + 40)\*\Nminus\Bigg)\*\Ss(1,1)+
\nonumber\\&&
(-28\*\Nminus+16-28\*\Nplus)\*\Ss(1,3)
+(-16-32\*\Nplus+32\*\Nminus)\*\Ss(2,-2)
+(-16+24\*\Nminus + 4\*\Nplus)\*\Ss(2,1)+
\nonumber\\&&
(36\*\Nminus+36\*\Nplus)\*\Ss(2,2)
+(16\*\Nminus-24\*\Nplus)\*\Ss(3,1)
+(8+20\*\Nminus+8\*\Nplus)\*\Sss(1,1,1)
+(24\*\Nminus+
\nonumber\\&&
24\*\Nplus)\*\Sss(1,1,2)
+(16\*\Nplus+16\*\Nminus)\*\Sss(1,2,1)
+(16\*\Nplus+24\*\Nminus)\*\Sss(2,1,1)
+(24\*\Nminus+24\*\Nplus)\*\Ssss(1,1,1,1)
\Bigg\} +
\nonumber\\&&
\colour4colour{\cf\*\ca}\*\Bigg\{8\*\S(-2)+
\Bigg(
\Bigg({1225\over 54}-36\*\z3
-{22\over 3}\*\z2\Bigg)\*\Nplus
-{24\over 5}\*\Nplustwo
+\Bigg({1580\over 27}-36\*\z3
-{22\over 3}\*\z2\Bigg)\*\Nminus +
\nonumber\\&&
24\*\z3-{24\over 5}\*\Nminustwo
-{593\over 45}\Bigg)\*\S(1)
+\Bigg(\Bigg({119\over 5}-8\*\z2\Bigg)\*\Nplus+{24\over 5}\*\Nplustwo
+\Bigg({863\over 15}-8\*\z2\Bigg)\*\Nminus-{24\over 5}\*\Nminustwo-
\nonumber\\&&
{38\over 3}\Bigg)\*\S(2) +
\Bigg({23\over 3}\*\Nplus+{24\over 5}\*\Nplustwo
-{24\over 5}\*\Nplusthree-4
+{11\over 3}\*\Nminus\Bigg)\*\S(3)+36\*\Nplus\*\S(4)
-8\*\Ss(-3,1)-24\*\Ss(-2,-2) +
\nonumber\\&&
(8+8\*\Nminus+8\*\Nplus)\*\Ss(1,-3) +
\Bigg( 4\*\Nplus-{24\over 5}\*\Nminusthree
-{24\over 5}\*\Nplusthree-8+{24\over 5}\*\Nplustwo
+{24\over 5}\*\Nminustwo +
\nonumber\\&&
4\*\Nminus \Bigg)\*\Ss(1,-2)
+\Bigg(\Bigg({311\over 9}+8\*\z2\Bigg)\*\Nplus- {112\over 3} +
\Bigg({392\over 9}+8\*\z2\Bigg)\*\Nminus\Bigg)\*\Ss(1,1)
+\Bigg(12\*\Nminus+12\*\Nplus-
\nonumber\\&&
8\Bigg)\*\Ss(1,3)
+ (8 -16\*\Nminus + 16\*\Nplus )\*\Ss(2,-2)
+\Bigg({22\over 3}\*\Nminus+{22\over 3}\*\Nplus\Bigg)\*\Ss(2,1)
+(4\*\Nminus+4\*\Nplus)\*\Ss(2,2)+
\nonumber\\&&
(-4\*\Nminus+4\*\Nplus)\*\Ss(3,1)
+(8\*\Nminus + 8\*\Nplus)\*\Sss(1,-2,1)
+(-16+24\*\Nplus+24\*\Nminus)\*\Sss(1,1,-2)
+\Bigg({22\over 3}\*\Nminus+
\nonumber\\&&
{22\over 3}\*\Nplus\Bigg)\*\Sss(1,1,1)
-(4\*\Nminus+4\*\Nplus)\*\Sss(1,1,2) +
(4\*\Nminus+4\*\Nplus)\*\Sss(1,2,1)
-36\*\S(-4) +7\*\z2+10\*\z3-{5465\over 72}
\Bigg\} +
\nonumber\\&&
\colour4colour{\cf\*\nf}\*\Bigg\{
\Bigg(\Bigg({4\over 3}\*\z2-{17\over 27}\Bigg)\*\Nplus-{14\over 9}+
\Bigg(-{188\over 27}+{4\over 3}\*\z2\Bigg)\*\Nminus\Bigg)\*\S(1)
+\Bigg({8\over 3}-{22\over 3}\*\Nminus-6\*\Nplus\Bigg)\*\S(2)  +
\nonumber\\&&
\Bigg(-{2\over 3}\*\Nplus-{2\over 3}\*\Nminus\Bigg)\*\S(3) +
\Bigg(-{26\over 9}\*\Nplus-{44\over 9}\*\Nminus+{4\over 3}\Bigg)\*\Ss(1,1)+
\Bigg(-{4\over 3}\*\Nplus-{4\over 3}\*\Nminus\Bigg)\*\Ss(2,1)+
\nonumber\\&&
\Bigg(-{4\over 3}\*\Nplus-{4\over 3}\*\Nminus\Bigg)\*\Sss(1,1,1)
-2\*\z2+{457\over 36}+8\*\z3
\Bigg\}
\Bigg]
\: \: ,\nonumber\\[2ex]
\label{eq:cTeqps2}
\lefteqn{ c_{T,\rm ps}^{(2)}(N) \, = \,}
\\&&
\delta(N-2)\*\colour4colour{\cf\*\nf}\*\Bigg[{2462\over 81}-{16\over 9}\*\z2\Bigg] +
\nonumber\\&&
\theta(N-3)\*\colour4colour{\cf\*\nf}\*
\Bigg[\Bigg(\Bigg({118\over 27}-{4\over 3}\*\z2\Bigg)\*\Nplus+
\Bigg({512\over 27}+{16\over 3}\*\z2\Bigg)\*\Nplustwo+
\Bigg( {982\over 27} -{4\over 3}\*\z2 \Bigg)\*\Nminus +
\Bigg({16\over 3}\*\z2+
\nonumber\\&&
{80\over 27}\Bigg)\*\Nminustwo-
{188\over 3}-8\*\z2\Bigg)\*\S(1)+
\Bigg(\Bigg(8\*\z2+{184\over 9}\Bigg)\*\Nplus+{128\over 9}\*\Nplustwo+
\Bigg(-{184\over 3}-8\*\z2\Bigg)\*\Nminus+32-
\nonumber\\&&
{16\over 3}\*\Nminustwo\Bigg)\*\S(2)+\Bigg(-{64\over 3}\*\Nminustwo-
{58\over 3}\*\Nplus+{16\over 3}\*\Nplustwo+{106\over 3}\*\Nminus\Bigg)\*\S(3)
+(-44\*\Nplus+44\*\Nminus)\*\S(4) +
\nonumber\\&&
\Bigg( {64\over 3}\*\Nplus -32
-{16\over 3}\*\Nminustwo-{16\over 3}\*\Nplustwo
+{64\over 3}\*\Nminus\Bigg)\*\Ss(1,-2)+
\Bigg({160\over 3}-28\*\Nminus-{8\over 3}\*\Nminustwo-12\*\Nplus -
\nonumber\\&&
{32\over 3}\*\Nplustwo \Bigg)\*\Ss(1,1) +
\Bigg(16-{16\over 3}\*\Nplustwo-{32\over 3}\*\Nminustwo
-{44\over 3}\*\Nplus+{44\over 3}\*\Nminus\Bigg)\*\Ss(2,1)
+(24\*\Nminus-24\*\Nplus)\*\Ss(3,1) +
\nonumber\\&&
\Bigg(-{16\over 3}\*\Nplustwo+8-{16\over 3}\*\Nminustwo
+{4\over 3}\*\Nminus+{4\over 3}\*\Nplus\Bigg)\*\Sss(1,1,1)
+(-8\*\Nplus+8\*\Nminus)\*\Sss(2,1,1)
\Bigg]
\: \: ,\nonumber\\[2ex]
\label{eq:cTeg2}
\lefteqn{ c_{T,\rm g}^{(2)}(N) \, = \,}
\\&&
\delta(N-2)\*\Bigg[
\colour4colour{\cf^2}\*\Bigg\{ {1408\over 15}\*\z3+{1120\over 9}\*\z2 -{140657\over 405} \Bigg\} +
\colour4colour{\cf\*\ca}\*\Bigg\{{104\over 9}\*\z2+48\*\z3-{26431\over 81}\Bigg\}
\Bigg] +
\nonumber\\&&
\theta(N-3)\*\Bigg[
\colour4colour{\cf^2}\*\Bigg\{\Bigg((36\*\z2+34-16\*\z3)\*\Nplus
-{16\over 5}\*\Nplustwo
+(184+32\*\z3+208\*\z2)\*\Nminus -
\nonumber\\&&
\Bigg({316\over 5}+96\*\z2+64\*\z3\Bigg)\*\Nminustwo
-148\*\z2+48\*\z3-{758\over 5}\Bigg)\*\S(1) +
\Bigg(\Bigg({246\over 5}+24\*\z2\Bigg)\*\Nplus+{16\over 5}\*\Nplustwo +
\nonumber\\&&
\Bigg(112\*\z2+{274\over 5}\Bigg)\*\Nminus
+\Bigg({144\over 5}-64\*\z2\Bigg)\*\Nminustwo-72\*\z2-136\Bigg)\*\S(2)
+ \Bigg(166\*\Nplus-230+{16\over 5}\*\Nplustwo-
\nonumber\\&&
{16\over 5}\*\Nplusthree + 64\*\Nminus\Bigg)\*\S(3)
+(-132+88\*\Nminus+44\*\Nplus)\*\S(4)
+(96-96\*\Nminus+32\*\Nminustwo-32\*\Nplus)\*\Ss(1,-3) +
\nonumber\\&&
\Bigg( 64\*\Nminus-{16\over 5}\*\Nplusthree-160
-{64\over 5}\*\Nminustwo+{64\over 5}\*\Nminusthree+96\*\Nplus
+{16\over 5}\*\Nplustwo\Bigg)\*\Ss(1,-2)
+\Bigg((16\*\z2+24)\*\Nplus +
\nonumber\\&&
(-8+64\*\z2)\*\Nminus+(8-32\*\z2)\*\Nminustwo-24-48\*\z2\Bigg)\*\Ss(1,1)
+(-16+16\*\Nplus)\*\Ss(1,2)
+\Bigg(-96\*\Nminustwo +
\nonumber\\&&
64\*\Nplus - 192+224\*\Nminus\Bigg)\*\Ss(1,3)
+(96\*\Nminustwo-12\*\Nplus+92-176\*\Nminus)\*\Ss(2,1)
+\Bigg(-128\*\Nminus +
\nonumber\\&&
128\Bigg)\*\Ss(2,-2)
+(64\*\Nminustwo - 128\*\Nminus+96-32\*\Nplus)\*\Ss(2,2)
+(-8\*\Nplus-80\*\Nminus+64\*\Nminustwo+24)\*\Ss(3,1) +
\nonumber\\&&
(-64\*\Nminustwo +64\*\Nplus - 192+192\*\Nminus )\*\Sss(1,1,-2)
+(-28\*\Nplus-208\*\Nminus+96\*\Nminustwo+140)\*\Sss(1,1,1) +
\nonumber\\&&
(-32\*\Nminus + 16\*\Nminustwo - 8\*\Nplus+24 )\*\Sss(1,1,2)
+(72-24\*\Nplus-96\*
\Nminus+48\*\Nminustwo)\*\Sss(1,2,1) +
\nonumber\\&&
(120+96\*\Nminustwo-40\*\Nplus - 176\*\Nminus )\*\Sss(2,1,1)
+ (80\*\Nminustwo-40\*\Nplus+120-160\*\Nminus)\*\Ssss(1,1,1,1)\Bigg\} +
\nonumber\\&&
\colour4colour{\ca\*\cf}\*\Bigg\{
\Bigg(\Bigg(-{1934\over 27}+72\*\z3+{8\over 3}\*\z2\Bigg)\*\Nplus+
\Bigg(-{928\over 27}-{32\over 3}\*\z2\Bigg)\*\Nplustwo+
\Bigg({440\over 3}\*\z2+240\*\z3 +
\nonumber\\&&
{5410\over 27}\Bigg)\*\Nminus +
\Bigg(-{248\over 3}\*\z2-96\*\z3-{4438\over 27}\Bigg)\*\Nminustwo
+70-216\*\z3-56\*\z2\Bigg)\*\S(1)
+ \Bigg(\Bigg(-{260\over 9} -
\nonumber\\&&
48\*\z2\Bigg)\*\Nplus-{256\over 9}\*\Nplustwo
+ (92-64\*\z2)\*\Nminus
+\Bigg({496\over 3}+64\*\z2\Bigg)\*\Nminustwo-200+48\*\z2\Bigg)\*\S(2) +
\nonumber\\&&
\Bigg({464\over 3}\*\Nminustwo -
{368\over 3}\*\Nminus+{44\over 3}\*\Nplus
-{32\over 3}\*\Nplustwo-36\Bigg)\*\S(3)
+(248\*\Nplus-320\*\Nminustwo+144\*\Nminus-72)\*\S(4) +
\nonumber\\&&
\Bigg(48\*\Nplus -
144 + 176\*\Nminus-80\*\Nminustwo\Bigg)\*\Ss(1,-3)
+ \Bigg({32\over 3}\*\Nplustwo
-{320\over 3}\*\Nminus+{80\over 3}\*\Nminustwo
-{200\over 3}\*\Nplus +
\nonumber\\&&
136\Bigg)\*\Ss(1,-2) +
\Bigg(\Bigg(32\*\z2-{28\over 3}\Bigg)\*\Nplus
+{32\over 3}\*\Nplustwo
+\Bigg(128\*\z2-{592\over 3}\Bigg)\*\Nminus
+\Bigg({356\over 3}-64\*\z2\Bigg)\*\Nminustwo +
\nonumber\\&&
{232\over 3}
-96\*\z2\Bigg)\*\Ss(1,1)
+(72-128\*\Nminus+64\*\Nminustwo-8\*\Nplus)\*\Ss(1,2)
+ \Bigg(96-32\*\Nplus-112\*\Nminus +
\nonumber\\&&
48\*\Nminustwo \Bigg)\*\Ss(1,3)
+(128\*\Nminus - 64\*\Nminustwo+32\*\Nplus-96)\*\Ss(2,-2)
+\Bigg({32\over 3}\*\Nplustwo+{496\over 3}\*\Nminustwo
+{16\over 3}\*\Nplus -
\nonumber\\&&
{784\over 3}\*\Nminus + 80\Bigg)\*\Ss(2,1)
+(96\*\Nminustwo-48\*\Nplus+144-192\*\Nminus)\*\Ss(2,2)
+\Bigg(32-96\*\Nminustwo-32\*\Nminus +
\nonumber\\&&
96\*\Nplus \Bigg)\*\Ss(3,1)
+\Bigg( 64\*\Nminus -48
-32\*\Nminustwo+16\*\Nplus\Bigg)\*\Sss(1,-2,1)
+(-32\*\Nminus-16\*\Nplus+48)\*\Sss(1,1,-2) +
\nonumber\\&&
\Bigg( {344\over 3}\*\Nminustwo -{80\over 3}\*\Nplus-{632\over 3}\*\Nminus  +
112+{32\over 3}\*\Nplustwo\Bigg)\*\Sss(1,1,1)
+ \Bigg( 64\*\Nminustwo-128\*\Nminus+96 -
\nonumber\\&&
32\*\Nplus \Bigg)\*\Sss(1,1,2)
+ ( 32\*\Nminustwo -16\*\Nplus - 64\*\Nminus +
48 )\*\Sss(1,2,1) +
\nonumber\\&&
(24\*\Nplus+24-32\*\Nminus-16\*\Nminustwo)\*\Sss(2,1,1)
+(24-32\*\Nminus+16\*\Nminustwo-8\*\Nplus)\*\Ssss(1,1,1,1)
\Bigg\}
\Bigg]
\: \: ,\nonumber\\[2ex]
\label{eq:cLeq2}
\lefteqn{ c_{L,\rm ns}^{(2)}(N) \, = \,}
\\&&
\delta(N-2)\*\Bigg[
\colour4colour{\cf^2}\*\Bigg\{ {48\over 5}\*\z3+{33\over 10} \Bigg\}
+\colour4colour{\cf\*\ca}\*\Bigg\{ -{24\over 5}\*\z3+{221\over 10} \Bigg\}
-{11\over 3}\*\colour4colour{\cf\*\nf} \Bigg] +
\nonumber\\&&
\theta(N-3)\*\Bigg[ \colour4colour{\cf^2}\*\Bigg\{
\Bigg(-10\*\Nplus+{32\over 5}\*\Nplustwo
+(-48\*\z3+39)\*\Nminus
-{129\over 5}-{48\over 5}\*\Nminustwo+48\*\z3\Bigg)\*\S(1)+
\nonumber\\&&
\Bigg( {8\over 5}\*\Nplus-{32\over 5}\*\Nplustwo+{82\over 5}\*\Nminus
-{48\over 5}\*\Nminustwo -2 \Bigg)\*\S(2)
+\Bigg( 12\*\Nminus+32\*\Nplus-{32\over 5}\*\Nplustwo
+{32\over 5}\*\Nplusthree -
\nonumber\\&&
44 \Bigg)\*\S(3) + (16-16\*\Nminus)\*\Ss(1,-3)
+\Bigg( 16\*\Nminus+{48\over 5}\*\Nminustwo+32\*\Nplus
-{32\over 5}\*\Nplustwo+{32\over 5}\*\Nplusthree
-{48\over 5}\*\Nminusthree -
\nonumber\\&&
48 \Bigg)\*\Ss(1,-2)
+(14\*\Nminus-10-4\*\Nplus)\*\Ss(1,1)
+(16-16\*\Nminus)\*\Ss(1,2)
+(16\*\Nminus-16)\*\Ss(1,3)+
\nonumber\\&&
(32-32\*\Nminus)\*\Ss(2,-2) +
(4-4\*\Nminus)\*\Ss(2,1)
+(-32+32\*\Nminus)\*\Sss(1,1,-2)
+(-8\*\Nminus+8)\*\Sss(1,1,1)
\Bigg\}+
\nonumber\\&&
\colour4colour{\cf\*\ca}\*\Bigg\{
\Bigg(-{10\over 3}\*\Nplus-{16\over 5}\*\Nplustwo
+\Bigg( 24\*\z3-{389\over 9} \Bigg)\*\Nminus
+{2023\over 45}-24\*\z3+{24\over 5}\*\Nminustwo\Bigg)\*\S(1)
+\Bigg({34\over 3}+
\nonumber\\&&
{24\over 5}\*\Nminustwo
-{24\over 5}\*\Nplus+{16\over 5}\*\Nplustwo
-{218\over 15}\*\Nminus\Bigg)\*\S(2)
+\Bigg(16+{16\over 5}\*\Nplustwo
-{16\over 5}\*\Nplusthree-16\*\Nplus\Bigg)\*\S(3) +
\nonumber\\&&
(8\*\Nminus-8)\*\Ss(1,-3) +
\Bigg(-{24\over 5}\*\Nminustwo-{16\over 5}\*\Nplusthree+{16\over 5}\*\Nplustwo+
24-8\*\Nminus+{24\over 5}\*\Nminusthree-16\*\Nplus\Bigg)\*\Ss(1,-2)+
\nonumber\\&&
\Bigg( {46\over 3}-{46\over 3}\*\Nminus \Bigg)\*\Ss(1,1)
+ (-8\*\Nminus+8)\*\Ss(1,3)+(16\*\Nminus-16)\*\Ss(2,-2)
+(16-16\*\Nminus)\*\Sss(1,1,-2)
\Bigg\} +
\nonumber\\&&
\colour4colour{\cf\*\nf}\*\Bigg\{
\Bigg( {50\over 9}\*\Nminus-{62\over 9}+{4\over 3}\*\Nplus \Bigg)\*\S(1) +
\Bigg( {4\over 3}\*\Nminus-{4\over 3}\Bigg)\*\S(2) +
\Bigg( {4\over 3}\*\Nminus-{4\over 3} \Bigg)\*\Ss(1,1)
\Bigg\}
\Bigg]
\: \: ,\nonumber\\[2ex]
\label{eq:cLeqps2}
\lefteqn{ c_{L,\rm ps}^{(2)}(N) \, = \,}
\\&&
-{{26\over 3}}\*\delta(N-2)\*\colour4colour{\cf\*\nf} +
\theta(N-3)\*\colour4colour{\cf\*\nf}\*\Bigg[
\Bigg( {32\over 3}\*\Nminus
- {160\over 3} +8\*\Nminustwo+{128\over 3}\*\Nplus-8\*\Nplustwo\Bigg)\*\S(1)+
\nonumber\\&&
\Bigg( {56\over 3}\*\Nplus - {8\over 3}\*\Nplustwo-{80\over 3}\*\Nminus +
{32\over 3}\*\Nminustwo \Bigg)\*\S(2)
+(24-24\*\Nminus)\*\S(3)+
\nonumber\\&&
\Bigg( {32\over 3}\*\Nplus - {16\over 3}\*\Nminus
- {8\over 3}\*\Nplustwo-8+{16\over 3}\*\Nminustwo \Bigg)\*\Ss(1,1)
+(8-8\*\Nminus)\*\Ss(2,1)
\Bigg]
\: \: ,\nonumber\\[2ex]
\label{eq:cLeg2}
\lefteqn{ c_{L,\rm g}^{(2)}(N) \, = \,}
\\&&
\delta(N-2)\*\Bigg[
\colour4colour{\cf^2}\*\Bigg\{ -{96\over 5}\*\z3-{108\over 5} \Bigg\}
+{272\over 3}\*\colour4colour{\cf\*\ca} \Bigg] +
\nonumber\\&&
\theta(N-3)\*\Bigg[
\colour4colour{\cf^2}\*\Bigg\{
\Bigg({56\over 3}\*\Nplus-{176\over 15}+{96\over 5}\*\Nminustwo
-{32\over 15}\*\Nplustwo-24\*\Nminus\Bigg)\*\S(1)
+\Bigg({64\over 5}\*\Nplus -{40\over 3}-
\nonumber\\&&
{104\over 5}\*\Nminus
+{96\over 5}\*\Nminustwo+{32\over 15}\*\Nplustwo\Bigg)\*\S(2)
+\Bigg(48-{32\over 15}\*\Nplusthree-48\*\Nminus
+{32\over 15}\*\Nplustwo\Bigg)\*\S(3) +
\nonumber\\&&
\Bigg( {32\over 15}\*\Nplustwo +
{32\over 3}\*\Nminus
+{64\over 5}\*\Nminustwo-{64\over 5}\*\Nminusthree
-{32\over 15}\*\Nplusthree -{32\over 3} \Bigg)\*\Ss(1,-2)
+(16-16\*\Nminus)\*\Ss(2,1) +
\nonumber\\&&
(8\*\Nplus-56\*\Nminus+32\*\Nminustwo+16)\*\Ss(1,1)
\Bigg\} +
\nonumber\\&&
\colour4colour{\cf\*\ca}\*\Bigg\{
\Bigg(-64\*\Nplus+{32\over 3}\*\Nplustwo-{160\over 3}
+256\*\Nminus-{448\over 3}\*\Nminustwo\Bigg)\*\S(1)
+\Bigg({16\over 3}\*\Nplustwo -{112\over 3}\*\Nplus -80 +
\nonumber\\&&
{688\over 3}\*\Nminus-{352\over 3}\*\Nminustwo\Bigg)\*\S(2)
+(-96+128\*\Nminustwo-32\*\Nminus)\*\S(3)
+(-64\*\Nminus+32\*\Nminustwo+32)\*\Ss(1,-2) +
\nonumber\\&&
(-64+128\*\Nminus-64\*\Nminustwo)\*\Ss(1,2)
+(64\*\Nminus-64)\*\Ss(2,1)
+\Bigg({896\over 3}\*\Nminus-128+{16\over 3}\*\Nplustwo  -
\nonumber\\&&
{464\over 3}\*\Nminustwo-{64\over 3}\*\Nplus\Bigg)\*\Ss(1,1)
+(-32\*\Nminustwo+64\*\Nminus-32)\*\Sss(1,1,1)
\Bigg\}
\Bigg]
\: \: ,\nonumber\\[2ex]
\label{eq:cAeq2}
\lefteqn{ c_{A,\rm ns}^{(2)}(N) \, = \,}
\\&&
\delta(N-1)\*\colour4colour{\cf}\*\Bigg[{11\over 3}\*\colour4colour{\ca}
-{2\over 3}\*\colour4colour{\nf}\Bigg]\*(-12)\*\z3 +
\nonumber\\&&
\theta(N-2)\*\Bigg[
\colour4colour{\cf^2}\*\Bigg\{ \Bigg(\Bigg(
{19\over 2}-40\*\z3-12\*\z2 \Bigg)\*\Nplus
+(-16-40\*\z3-12\*\z2)\*\Nminus-19+48\*\z3\Bigg)\*\S(1) +
\nonumber\\&&
((12\*\z2-92)\*\Nplus+78+(4\*\z2-92)\*\Nminus)\*\S(2)
-(94\*\Nminus+66\*\Nplus)\*\S(4)
+\Bigg( 20\*\Nminus-90\*\Nplus +
\nonumber\\&&
16\*\Nplustwo-12 \Bigg)\*\S(3)
+(24+12\*\Nminus-12\*\Nplus)\*\Ss(1,2)
+(-12\*\Nminus-12\*\Nplus-16)\*\Ss(1,3)
+\Bigg(16-
\nonumber\\&&
32\*\Nminus \Bigg)\*\Ss(2,-2)
+(-8\*\Nplus+44\*\Nminus-24)\*\Ss(2,1)
+(36\*\Nminus+36\*\Nplus)\*\Ss(2,2)
+\Bigg(-32\*\Nminus+16 -
\nonumber\\&&
32\*\Nplus \Bigg)\*\Ss(1,-3)
+(16\*\Nminustwo
-24\*\Nplus+16\*\Nplustwo
-24\*\Nminus+16)\*\Ss(1,-2)
+\Bigg((-32\*\z2-33)\*\Nplus+
\nonumber\\&&
28
+(-22-32\*\z2)\*\Nminus\Bigg)\*\Ss(1,1)
+(24\*\Nminus+16\*\Nplus)\*\Sss(2,1,1)
-(16\*\Nminus+16\*\Nplus)\*\Sss(1,-2,1) +
\nonumber\\&&
(-32 -16\*\Nminus
-16\*\Nplus )\*\Sss(1,1,-2)
+(28\*\Nminus
+16\*\Nplus -8 )\*\Sss(1,1,1)
+(24\*\Nplus+24\*\Nminus)\*\Sss(1,1,2)+
\nonumber\\&&
(16\*\Nplus+16\*\Nminus)\*\Sss(1,2,1)
+(24\*\Nplus+24\*\Nminus)\*\Ssss(1,1,1,1)
+72\*\S(-4)
-16\*\S(-2)
+16\*\Ss(-3,1)+
\nonumber\\&&
48\*\Ss(-2,-2)
-8\*\Nplus\*\Ss(3,1) + {331\over 8} - 12\*\z3+59\*\z2
\Bigg\} +
\nonumber\\&&
\colour4colour{\cf\*\ca}\*\Bigg\{ \Bigg(\Bigg(-{22\over 3}\*\z2+{895\over 54}
-12\*\z3\Bigg)\*\Nplus + \Bigg(-{22\over 3}\*\z2+{1415\over 27}
-12\*\z3\Bigg)\*\Nminus-{95\over 9} - 24\*\z3\Bigg)\*\S(1) +
\nonumber\\&&
\Bigg(\Bigg({193\over 3}-8\*\z2\Bigg)\*\Nplus-{250\over 3}
+\Bigg({263\over 3}-8\*\z2\Bigg)\*\Nminus\Bigg)\*\S(2)
+\Bigg(-4-{37\over 3}\*\Nminus+{95\over 3}\*\Nplus-8\*\Nplustwo\Bigg)\*\S(3) +
\nonumber\\&&
(8+4\*\Nminus+4\*\Nplus)\*\Ss(1,3)
+(16\*\Nminus-8)\*\Ss(2,-2)
+\Bigg(-{2\over 3}\*\Nminus+{46\over 3}\*\Nplus \Bigg)\*\Ss(2,1)
+(4\*\Nplus+4\*\Nminus)\*\Ss(2,2) +
\nonumber\\&&
36\*\Nminus\*\S(4)
+(4\*\Nminus-4\*\Nplus)\*\Ss(3,1)
-8\*\Ss(-3,1)
-24\*\Ss(-2,-2)
+(16\*\Nminus+16\*\Nplus -8 )\*\Ss(1,-3)
+\Bigg(  12\*\Nminus +
\nonumber\\&&
12\*\Nplus -8\*\Nminustwo -8\*\Nplustwo-8 \Bigg)\*\Ss(1,-2)
+\Bigg(\Bigg(8\*\z2+{305\over 9}\Bigg)\*\Nplus
-36+\Bigg(8\*\z2+{386\over 9}\Bigg)\*\Nminus\Bigg)\*\Ss(1,1) +
\nonumber\\&&
(8\*\Nminus+8\*\Nplus)\*\Sss(1,-2,1)
+(16+8\*\Nminus+8\*\Nplus)\*\Sss(1,1,-2)
+\Bigg({22\over 3}\*\Nplus+{22\over 3}\*\Nminus \Bigg)\*\Sss(1,1,1)
-36\*\S(-4) +
\nonumber\\&&
8\*\S(-2)
+(-4\*\Nplus-4\*\Nminus)\*\Sss(1,1,2)
+(4\*\Nplus+4\*\Nminus)\*\Sss(1,2,1)
-{5465\over 72}+10\*\z3+7\*\z2
\Bigg\} +
\nonumber\\&&
\colour4colour{\cf\*\nf}\*\Bigg\{
\Bigg( \Bigg( {4\over 3}\*\z2-{131\over 27}\Bigg)\*\Nplus
+{62\over 9}+\Bigg({4\over 3}\*\z2-{302\over 27}\Bigg)\*\Nminus\Bigg)\*\S(1)
+\Bigg({16\over 3}-{22\over 3}\*\Nplus-{26\over 3}\*\Nminus\Bigg)\*\S(2) +
\nonumber\\&&
\Bigg(-{2\over 3}\*\Nplus-{2\over 3}\*\Nminus \Bigg)\*\S(3)
+\Bigg(-{4\over 3}\*\Nplus-{4\over 3}\*\Nminus \Bigg)\*\Ss(2,1)
+\Bigg(-{38\over 9}\*\Nplus+4-{56\over 9}\*\Nminus \Bigg)\*\Ss(1,1) +
\nonumber\\&&
\Bigg(-{4\over 3}\*\Nplus-{4\over 3}\*\Nminus \Bigg)\*\Sss(1,1,1)
+{457\over 36}+8\*\z3-2\*\z2
\Bigg\}
\Bigg]
\: \: .\nonumber
\end{eqnarray}

\renewcommand{\theequation}{C.\arabic{equation}}
\setcounter{equation}{0}
%
%
\section*{Appendix C: The exact $x$-space results}
\label{sec:appendix-C}
Here we write down the full $x$-space results of the
time-like splitting functions and the coefficient functions
$c_{T}$, $c_{L}$ and $c_{A}$ up to second order in $\ar = \as /
(4\pi)$, expressed in terms of harmonic polylogarithms in the
notation $H_{m_1,...,m_w}(x)$, $m_j = 0,\pm 1$ of
Ref.~\cite{Remiddi:1999ew}. Below we use the short-hand notation
\begin{equation}
  \label{eq:HPL}
  H_{{\footnotesize \underbrace{0,\ldots ,0}_{\scriptstyle m} },\,
  \pm 1,\, {\footnotesize \underbrace{0,\ldots ,0}_{\scriptstyle n} },
  \, \pm 1,\, \ldots}(x) \, = \, H_{\pm (m+1),\,\pm (n+1),\, \ldots}(x)
\, ,
\end{equation}
suppress the argument $x$ for brevity and define
\begin{eqnarray}
  \label{eq:pqq}
  p_{\rm{qq}}(x) &\! =\! & 2\, (1 - x)^{-1} - 1 - x \, ,
\\
  p_{\rm{gg}}(x) &\! =\! & (1-x)^{-1} + x^{\,-1} - 2 + x - x^{\,2} \, .
\nonumber
\end{eqnarray}
All divergences for $x \to 1 $ are understood in the sense of
$+$-distributions.

The well known results at LO are:
\begin{eqnarray}
&& P^{(0)}_{\rm qq}(x) \, = \,
\colour4colour{\cf}\*\left(2\*\pqq(x) + 3\*\delta(1 - x)\right)
\: \: ,
\\[2ex]
&& P^{(0)}_{\rm qg}(x) \, = \,
\colour4colour{\nf}\*\left(2 - 4\*x + 4\*x^2\right)
\: \: ,
\\[2ex]
&& P^{(0)}_{\rm gq}(x) \, = \,
\colour4colour{\cf}\*\left( - 4 + {4\over x} + 2\*x\right)
\: \: ,
\\[2ex]
&& P^{(0)}_{\rm gg}(x) \, = \,
\colour4colour{\ca}\*\left(4\*\pgg(x) + {11\over 3}\*\delta(1-x) \right) -
\colour4colour{\nf}\*{2\over 3}\*\delta(1-x)
\: \: .
\end{eqnarray}

The NLO splitting functions read\footnote{
We quote $P_{\,\rm qg}^{\,(1)}$ and $P_{\,\rm gg}^{\,(1)}$ here for completeness
(see discussion in Section~\ref{sec:results}).}:
\begin{eqnarray}
\lefteqn{ P^{(1)+}_{\rm ns}(x) \, = \,}
\\&&
\colour4colour{\cf^2}\*\Bigg\{\pqq(x)\*(-8\*\H(2)+6\*\H(0)-16\*\Hh(0,0)-8\*\Hh(1,0))
+ \pqq(-x)\*(-8\*\z2+8\* \Hh(0,0)-16\*\Hh(-1,0)) +
\nonumber\\&&
(-6+2\*x)\*\H(0) + (4+4\*x)\*\Hh(0,0)-4+4\*x+\left(-12\*\z2
+{3\over 2}+24\*\z3\right)\*\delta(1-x)\Bigg\} +
\nonumber\\&&
\colour4colour{\ca\*\cf}\*\Bigg\{ \pqq(x)\*\left(4\*\Hh(0,0)-4\*\z2+{134\over
9}+{22\over 3}\*\H(0)\right) +
\pqq(-x)\*(-4\*\Hh(0,0)+8\*\Hh(-1,0)+4\*\z2) +
\nonumber\\&&
{56\over 3}\*(1-x) + \left({44\over 3}\*\z2+{17\over
6}-12\*\z3\right)\*\delta( 1-x) \Bigg\} +
\nonumber\\&&
\colour4colour{\nf\*\cf}\*\Bigg\{ \pqq(x)\*\left(-{4\over 3}\*\H(0)-{20\over 9}\right)
-{8\over 3}\*(1-x) + \left(-{1\over 3}
-{8\over 3}\*\z2\right)\*\delta(1-x)\Bigg\}
\: \: ,\nonumber\\[2ex]
\lefteqn{ P^{(1)-}_{\rm ns}(x) \, = \,}
\\&&
\colour4colour{\cf^2}\*\Bigg\{
\pqq(x)\*(6\*\H(0)-8\*\H(2)-8\*\Hh(1,0)-16\*\Hh(0,0)) +
\pqq(-x)\*(16\*\Hh(-1, 0)-8\*\Hh(0,0)+8\*\z2) +
\nonumber\\&&
(-14\*x-22)\*\H(0)+(4+4\*x)\*\Hh(0,0)+36\*x-36 +\left({3\over
2}-12\*\z2 +24\*\z3\right)\*\delta(1-x) \Bigg\} +
\nonumber\\&&
\colour4colour{\cf\*\ca}\*\Bigg\{ \pqq(x)\*\left(-4\*\z2+4\*\Hh(0,0)+{22\over
3}\*\H(0)+{134\over 9}\right) +
(-4\*\z2-8\*\Hh(-1,0)+4\*\Hh(0,0))\*\pqq(-x) +
\nonumber\\&&
(8\*x+8)\*\H(0)-{104\over 3}\*x+{104\over 3}
+\left(-12\*\z3+{17\over 6} +{44\over 3}\*\z2\right)\*\delta(1-x)
\Bigg\} +
\nonumber\\&&
\colour4colour{\cf\*\nf}\*\Bigg\{ \pqq(x)\*\left(-{4\over 3}\*\H(0)-{20\over
9}\right)+{8\over 3}\*x-{8\over 3} -\left({1\over 3}+{8\over
3}\*\z2\right)\*\delta(1-x) \Bigg\}
\: \: ,\nonumber\\[2ex]
\lefteqn{ P^{(1)\rm s}_{\rm qq}(x) \, = \,
\colour4colour{\nf\*\cf}\*\Bigg\{
8\*(1+x)\*\Hh(0,0)-\left({32\over 3}\*x^2+20+36\*x\right)\*\H(0)
+{{16\*(x-1)\*(14\*x^2+23\*x +5)}\over {9\*x}}\Bigg\}
\: \: , }
\\[2ex]
\lefteqn{ P^{(1)}_{\rm qg}(x) \, = \,}
\\&&
\colour4colour{\nf\*\cf}\*\Bigg\{(8\*x-10+8\*x^2)\*\H(0)+(8\*x^2+12-8\*x)\*\H(1) +
(8\*x-4-16\*x^2)\*\Hh(0,0) +
\nonumber\\&&
(-16\*x+16\*x^2+8)\*(\H(2) -3\*\Hh(1,0) -\Hh( 1,1) - \z2)
-40\*x^2-24+46\*x\Bigg\}+
\nonumber\\&&
\colour4colour{\nf\*\ca}\*\Bigg\{(-{136\over 3}\*x-{16\over 3}\*x^2-{16\over
3})\*\H( 0) + ({40\over 3}\*x-{40\over 3}\*x^2-{44\over 3})\*\H(1)
+
\nonumber\\&&
(-16\*x+16\*x^2+8)\*( \Hh(1,1) - 2\*\H(2) +2\*\Hh(1,0)) +
(-16\*x-8-16\*x^2)\*\Hh(-1,0)+
\nonumber\\&&
(8+48\*x)\*\Hh(0,0) - 16\*\z2\*x + {
{4\*(-95\*x^2-20+178\*x^3+13\*x)} \over {9\*x} }\Bigg\} +
\nonumber\\&&
\colour4colour{\nf^2}\*\Bigg\{\left({16\over 3}\*x^2+{8\over 3}-{16\over
3}\*x\right)\*(\H(1)-\H(0))-{40\over 9}+{32\over 9}\*x-{32\over
9}\*x^2\Bigg\}
\: \: ,\nonumber\\[2ex]
\lefteqn{ P^{(1)}_{\rm gq}(x) \, = \,}
\\&&
\colour4colour{\cf^2}\*\Bigg\{(-32+2\*x)\*\H(0)-8\*x\*\H(1) +{
{8\*(-2\*x+x^2+2)}\over x}\*(\Hh(1,1) +2\*\Hh(1,0) -2\*\H(2)) +
\nonumber\\&&
(8-4\*x)\*\Hh(0,0)+18\*x-2\Bigg\}+
\nonumber\\&&
\colour4colour{\ca\*\cf}\*\Bigg\{ { {4\*(27\*x^2+8\*x^3+24\*x-18)} \over {3\*x}
}\*\H(0)+8\*x\*\H(1)+{ {8\*(-2\*x+x^2+2)}\over x}\*( \H(2)
-3\*\Hh(1,0) - \Hh(1,1) ) +
\nonumber\\&&
{ {8\*(2\*x+x^2+2)}\over x}\*\Hh(-1,0)-{{8\*(2\*x+3\*x^2+4)}\over
x}\*\Hh(0,0) + 16\*\z2-{ {4\*(-45\*x+9\*x^2-17+44\*x^3)}\over
{9\*x}}\Bigg\}
\: \: ,\nonumber\\[2ex]
\lefteqn{ P^{(1)}_{\rm gg}(x) \, = \,}
\\&&
\colour4colour{\nf\*\cf}\*\Bigg\{ { {4\*(21\*x^2+8\*x^3+15\*x+8)}\over {3\*x}
}\*\H(0)+(8\*x+8)\*\Hh(0,0) -
\nonumber\\&&
{ {8\*(x-1)\*(41\*x^2+14\*x+23)}\over {9\*x}} -
2\*\delta(1-x)\Bigg\} +
\nonumber\\&&
\colour4colour{\ca^2}\*\Bigg\{ \pgg(x)\*\left({268\over
9}-16\*\H(2)-8\*\z2+{88\over 3}\*\H(0)-24\*
\Hh(0,0)-16\*\Hh(1,0)\right) +
\nonumber\\&&
(-8\*\z2+8\*\Hh(0,0)-16\*\Hh(-1,0))\*\pgg(-x)-{
{4\*(44-11\*x+25\*x^2)}\over {3\*x}}\*\H(0) +
(-32-32\*x)\*\Hh(0,0)+
\nonumber\\&&
{ {2\*(x-1)\*(134\*x^2-109\*x+134)}\over {9\*x}} +
\left(12\*\z3+{32\over 3}\right)\*\delta(1-x)\Bigg\} +
\nonumber\\&&
\colour4colour{\nf\*\ca}\*\Bigg\{\left(-{16\over 3}\*\H(0)-{40\over
9}\right)\*\pgg(x)-\left({8\over 3}+{8\over 3}\*x\right) \*\H(0) +
{ {4\*(x-1)\*(13\*x^2+4\*x+13)}\over {9\*x}} -{8\over
3}\*\delta(1-x) \Bigg\}
\: \: .\nonumber
\end{eqnarray}

The coefficient functions at order ${\cal O}(\as)$  read:
\begin{eqnarray}
&& c_{T,\rm q}^{(1)}(x) \, = \,
\colour4colour{\cf}\*\left( \pqq(x)\*\left(-{3\over
2}+4\*\H(0)-2\*\H(1)\right)-{9\over 2}\*x+{3\over 2}
+(8\*\z2-9)\*\delta(1-x) \right)
\: \: ,
\\[2ex]
&& c_{L,\rm q}^{(1)}(x) \, = \,
2\*\colour4colour{\cf}
\: \: ,
\\[2ex]
&& c_{A,\rm q}^{(1)}(x) - c_{T,\rm q}^{(1)}(x) \, = \,
\colour4colour{\cf}\*(2\*x-2)
\: \: ,
\\[2ex]
&& c_{T,\rm g}^{(1)}(x) \, = \,
\colour4colour{\cf}\*\left({ {8\*(-2\*x+2+x^2)}\over x} \*\H(0)
-{{4\*(-2\*x+2+x^2)}\over x}\*\H(1)+{{8\*(-1+x)}\over
x}\right)
\: \: ,
\\[2ex]
&& c_{L,\rm g}^{(1)}(x) \, = \,
\colour4colour{\cf}\*{{8\*(1-x)}\over x}
\: \: .
\end{eqnarray}

The coefficient functions at order ${\cal O}(\as^2)$ read:
\begin{eqnarray}
\lefteqn{ c_{T,\rm ns}^{(2)}(x) \, = \,}
\\&&
\colour4colour{\cf^2}\*\Bigg\{
\pqq(x)\*\Bigg((20\*\z2-53)\*\H(0)+6\*\z2+6\*\H(2)+{27\over
2}\*\H(1)+4\*\H(3)+12\*\Hh(1,0 )+{51\over
4}-18\*\Hh(1,1) +
\nonumber\\&&
33\*\Hh(0,0)-76\*\z3+20\*\Hhh(1,0,0)+24\*\Hhh(1,1,0)-24\*
\Hhh(1,1,1)+16\*\Hh(1,2)-36\*\Hh(2,0)+24\*\Hh(-2,0)+20\*\Hh(2,1)-
\nonumber\\&&
80\*\Hhh(0,0,0)\Bigg) + \pqq(-x)\*\Bigg(-32\*\H(-1)\*\z2+(-8\*\z2-8)\*\H(0)-8\*\Hh(-2,0)-8\*\H(
3)+36\*\Hhh(0,0,0)-
\nonumber\\&&
32\*\Hhh(-1,-1,0)-24\*\Hhh(-1,0,0) + 16\*\Hh(-1,2)+28\*\z3 \Bigg) + (-8\*x-8)\*\z2\*\H(-1)+
\Bigg((8\*x+8)\*\z2 +
\nonumber\\&&
{{-59\*x^2+48+48\*x^3+71\*x}\over {5\*x}}\Bigg)\*\H(0)
+ \left((-8+8\*x)\*\z2-{75\over 2}\*x+{53\over
2}\right)\*\H(1)+(18+2\*x)\*\H(2) -
\nonumber\\&&
(12+12\*x)\*\H(3) - {{8\*(x+1)\*(6\*x^4-6\*x^3+x^2-6\*x+6)}\over
{5\*x^2}}\*\Hh(-1,0)+\left(-33+3\*x+{48\over 5}\*x^
3\right)\*\Hh(0,0)+
\nonumber\\&&
(40\*x-16)\*\Hh(1,0) +16\*\Hh(-2,0) + (-10\*x-2)\*\Hh(1,1)+(4\*x+4)\*\Hh(2,1)+(-\
16-16\*x)\*\Hhh(-1,-1,0) +
\nonumber\\&&
(8\*x+8)\*\Hhh(-1,0,0)+(22\*x+22)\*\Hhh(0,0,0) + (-8\*x+8)\*\Hhh(1,0,0) +
\left(6\*x-{48\over 5}\*x^3-18\right)\*\z2 +
\nonumber\\&&
(8+24\*x)\*\z3+ {{3\*(271\*x-101\*x^2-64+64\*x^3)}\over {20\*x}} +
\left({331\over 8}+30\*\z2^2-78\*\z3-39\*\z2\right)\*\delta(1-x) \Bigg\}  +
\nonumber\\&&
\colour4colour{\cf\*\ca}\*\Bigg\{
\pqq(x)\*\left(
\left({103\over 3}
-16\*\z2\right)\*\H(0)
-\left( \ 4\*\z2
+{367\over 18}\right)\*\H(1)
-4\*\Hh(2,0)
-{3155\over 108}
-{22\over 3}\*\Hh(1,1)+14\*\z3 -  \right.
\nonumber\\&&
\left. 12\*\Hh(-2,0 )
-{11\over 3}\*\Hh(0,0)
-8\*\Hhh(1,0,0)
-4\*\Hhh(1,1,0)
+{22\over 3}\*\H(2)
+18\*\Hhh(0,0,0)
+4\*\Hh(1,2)\right)
+ \pqq(-x)\*\Bigg(16\*\H(-1)\*\z2 +
\nonumber\\&&
(4\*\z2+4)\*\H(0)-18\*\Hhh(0,0,0)+4\*\Hh(-2,0)-14\*\z3+
12\*\Hhh(-1,0,0)-8\*\Hh(-1,2)+16\*\Hhh(-1,-1,0)+4\*\H(3) \Bigg) +
\nonumber\\&&
(4\*x+4)\*\z2\*\H(-1)-{{2\*(-13\*x^2+36-108\*x+36\*x^3)}\over {15\*x}}\*\H(0)+\left((4-4\*x
)\*\z2-{139\over 6}+{85\over 6}\*x\right)\*\H(1)-
\nonumber\\&&
8\*\Hh(-2,0) + {{4\*(x+1)\*(6\*x^4-6\*x^3+x^2-6\*x+6)}\over {5\*x^2}} \*\Hh(-1,0)-{4\over
5}\*x\*(-5+6\*x^2)\*\Hh(0,0)+
\nonumber\\&&
(8\*x+8)\*\Hhh(-1,-1,0) + (-4-4\*x)\*\Hhh(-1,0,0)+(4\*x-4)\*\Hhh(1,0,0) + {4\over
5}\*x\*(-5+6\*x^2)\*\z2-8\*\z3\*x -
\nonumber\\&&
{{(2039\*x^2-864+864\*x^3+4411\*x)}\over {180\*x}} +\left(-{49\over 5}\*\z2^2-{5465\over
72}+{140\over 3}\*\z3+{215\over 3}\*\z2\right)\* \delta(1-x) \Bigg\} +
\nonumber\\&&
\colour4colour{\cf\*\nf}\*\Bigg\{
\pqq(x)\*\left(-{16\over 3}\*\H(0)+{4\over 3}\*\Hh(1,1)+{29\over
9}\*\H(1)+{247\over 54}-{4\over 3}\*\H(2)+{2\over 3}\*\Hh(0,0)\right)
+ \left(-2+{2\over 3}\*x\right)\*\H(0)+
\nonumber\\&&
\left({5\over 3} + {1\over 3}\*x\right)\*\H(1) +{43\over 18}+{71\over 18}\*x
+\left({4\over 3}\*\z3+{457\over 36}-{38\over 3}\*\z2\right)\*\delta(1-x)\Bigg\}
\: \: ,\nonumber\\[2ex]
\lefteqn{ c_{T,\rm ps}^{(2)}(x) \, = \,}
\\&&
\colour4colour{\cf\*\nf}\*\Bigg\{
\left((16+16\*x)\*\z2-{{8\*(16\*x^3+75\*x+6+39\*x^2)}\over
{9\*x}}\right)\*\H(0) - (24\*x+24)\*\H(3) -
\nonumber\\&&
{{4\*(x-1)\*(8\*x^2+ 25\*x+2)}\over
{3\*x}}\*\H(1)  + {{4\*(x+1)\*(4\*x^2+11\*x-8)}\over
{3\*x}}\*\H(2)-{{16\*( x+1)^3}\over {3\*x}}\*\Hh(-1,0)+
\nonumber\\&&
{{2\*(32-21\*x^2+8\*x^3-21\*x)}\over {3\*x}}\*\Hh(0,0) -
{{4\*(x-1)\*(4\*x^2+ 7\*x+4)}\over {3\*x}}\*\Hh(1,1)+(8+8\*x)\*\Hh(2,1)+
\nonumber\\&&
(44\*x+44)\*\Hhh(0,0,0)+(-8-32\*x)\*\z2 + (16+16\*x)\*\z3
+{{2\*(x-1)\*(256\*x^2+571\*x+40)}\over {27\*x}} \Bigg\}
\: \: ,\nonumber\\[2ex]
\lefteqn{ c_{T,\rm g}^{(2)}(x) \, = \,}
\\&&
\colour4colour{\cf^2}\*\Bigg\{
{{32\*(x+1)^2}\over x}\*\z2\*\H(-1)+\left((-16\*x+32)\*\z2-
{{2\*(-209\*x-72+131\*x^2+8\*x^3)}\over {5\*x}}\right)\*\H(0) +
\nonumber\\&&
\left(-{{8\*(5\*x^2-10\*x+6)}\over x}\*\z2
+ {{8\*( 3\*x^2-1)}\over x}\right)\*\H(1)
+{{4\*( 3\*x^2-20\*x+24)}\over x}\*\H(2)
-{{8\*(-2\*x+x^2+8)}\over x}\*\H(3) -
\nonumber\\&&
128\*\Hh(-2,0) + {{16\*(-20\*x^2-30\*x^3+x^5-4)} \over {5\*x^2}}\*\Hh(-1,0) +
\left(-64+166\*x-{16\over 5}\*x^3\right)\*\Hh(0,0)-16\*\Hh(1,0)\*x -
\nonumber\\&&
{{4\*(-28\*x+7\*x^ 2+24)}\over x}\*\Hh(1,1) +
{{24\*(-2\*x+2+x^2)}\over x}\*\Hh(1,2) - {{32\*(-2\*x+2+x^2)}\over x}\*\Hh(2,0) +
\nonumber\\&&
{{8\*(-10\*x+12+5\*x^2)}\over x}\*\Hh(2,1) + {{64\*(x+1)^2}\over
x}\*\Hhh(-1,-1,0)-{{32\*(x+1)^2}\over x}\*\Hhh(-1,0,0) +
(-44\*x+88)\*\Hhh(0,0,0) +
\nonumber\\&&
{{32\*(2\*x^2-4\*x+3)}\over x}\*\Hhh(1,0,0) +
{{8\*(-2 \*x+2+x^2)}\over x}\*\Hhh(1,1,0) - {{40\*(-2\*x+2+x^2)}\over x}\*\Hhh(1,1,1) +
\nonumber\\&&
\left(-32+{16\over 5}\*x^3-72\*x\right)\*\z2 - {{16\*(4+5\*x^2-2\*x)}\over x}\*\z3 -
{{2\*(302\*x+8\*x^3-77\*x^2-158)}\over {5\*x}}\Bigg\} +
\nonumber\\&&
\colour4colour{\cf\*\ca}\*\Bigg\{
\left(-{{16\*(8\*x+x^2-2)}\over x}\*\z2
+{{4\*(579\*x+372+64\*x^3+129\*x^2)}\over {9\*x}}\right)\*\H(0) +
\nonumber\\&&
{{8\*(4+x^2+2\*x)}\over x}\*\z2\*\H(-1)
+\left({{8\*(3\*x^2-6\*x+4)}\over x}\*\z2
+{{4\*(8\*x^3+x^2-89+59\*x)}\over {3\*x}}\right)\*\H(1) -
\nonumber\\&&
{{16\*(3\*x^2+18\*x-31+2\* x^3)}\over {3\*x}}\*\H(2)
+{{32\*(4\*x+3+3\*x^2)}\over x}\*\H(3)
+{{32\*(2\*x+2+x^2)}\over x}\*\Hh(-2,0)+
\nonumber\\&&
{{8\*(4\* x^3+10+30\*x+21\*x^2)}\over {3\*x}}\*\Hh(-1,0)
-{{16\*(2\*x+2+x^2)}\over x}\*\Hh(-1,2)
-{{4\*(24\*x-3\*x^2 +116+8\*x^3)}\over {3\*x}}\*\Hh(0,0) +
\nonumber\\&&
{{8\*(8+x^2-8\*x)}\over x}\*\Hh(1,0)
+{{8\*(36\*x-6\*x^2+4\*x^3-43)}\over {3\*x}}\*\Hh(1,1) +
{{16\*(-2\*x+2+x^2)}\over x}\*\Hh(1,2) -
\nonumber\\&&
{{48\*(-2\*x+2+x^2)}\over x}\*\Hh(2,0)
-{{8\*(3\*x^2+6\*x+2)}\over x}\*\Hh(2,1)
+ (-16\*x-32)\*\Hhh(-1,-1,0) +
\nonumber\\&&
{{16\*(5+6\*x+3\*x^2)}\over x}\*\Hhh(-1, 0,0)
-{{8\*(40+22\*x+31\*x^2)}\over x}\*\Hhh(0,0,0)
-{{16\*(2\*x^2-4\*x+3)}\over x}\*\Hhh(1,0,0) +
\nonumber\\&&
{{32\*(-2\*x+2+x^2)}\over x}\*\Hhh(1,1,0)
-{{8\*(-2\*x+2+x^2)}\over x}\*\Hhh(1,1,1)
+{{8\*(-7+8\*x^2+4\*x)}\over x}\*\z2 -
\nonumber\\&&
{{8\*(17\*x^2+30-8\*x)}\over x}\*\z3
-{{2\*(486\*x-2219+464\*x^3+1431\*x^2)}\over {27\*x}}\Bigg\}
\: \: ,\nonumber\\[2ex]
\lefteqn{ c_{L,\rm ns}^{(2)}(x) \, = \,}
\\&&
\colour4colour{\cf^2}\*\Bigg\{
16\*\H(-1)\*\z2+{{2\*(17\*x-24+12\*x^2+16\*x^3)}\over {5\*x}}\*\H(0) + (16\*\z2-4\*x-14)\* \H(1) - 4\*\H(2) -
\nonumber\\&&
{{16\*(2\*x^5+5\*x^2-3+10\*x^3)}\over {5\*x^2}}\*\Hh(-1,0)+
\left({32\over 5}\*x^3-12+32\*x\right)\*\Hh(0,0)-16\*\Hh(1,0)+8\*\Hh(1,1)-32\*\Hh(-2,0) +
\nonumber\\&&
32\*\Hhh(-1,-1,0)
- 16\*\Hhh(-1,0,0)
-16\*\Hhh(1,0,0)
+\left(4-{32\over 5}\*x^3-32\*x\right)\*\z2
+{{48-147\*x-18\*x^2+32\* x^3 }\over {5\*x}} \Bigg\} +
\nonumber\\&&
\colour4colour{\cf\*\ca}\*\Bigg\{
-8\*\H(-1)\*\z2-{{2\*(-12\*x^2-36+73\*x+24\*x^3)}\over {15\*x}}\*\H(0) +
\left({46\over 3}-8\*\z2\right)\*\H(1) +
\nonumber\\&&
{{8\*(2\*x^5+5\*x^2-3+10\*x^3)}\over {5\*x^2}}\*\Hh(-1,0) -
{16\over 5}\*x\*(x^2+5)\*\Hh(0,0)+16\*\Hh(-2,0)-16\*\Hhh(-1,-1,0)+8\*\Hhh(-1,0,0) +
\nonumber\\&&
8\*\Hhh(1,0,0) +
{16\over 5}\*x\*( x^2+5)\*\z2- {{144\*x^3+294\*x^2+216-1729\*x} \over {45\*x}} \Bigg\} +
\nonumber\\&&
\colour4colour{\cf\*\nf}\*\Bigg\{-{4\over 3}\*\H(1)-{50\over 9}+{4\over 3}\*\H( 0)+{4\over 3}\*x \Bigg\}
\: \: ,\nonumber\\[2ex]
\lefteqn{ c_{L,\rm ps}^{(2)}(x) \, = \,}
\\&&
\colour4colour{\cf\*\nf}\*\Bigg\{
{{8\*(-6\*x^2+x^3-6\*x+4)}\over {3\*x}}\*\H(0)- {{8\*(x-1)\*(x^2-2\*x-2)}\over {3\*x}}\*\H(1) -
8\*\H(2) +
\nonumber\\&&
24\*\Hh(0,0) + 8\*\z2-{{8\*(x-1)\*(3\*x^2-10\*x-3)}\over {3\*x}} \Bigg\}
\: \: ,\nonumber\\[2ex]
\lefteqn{ c_{L,\rm g}^{(2)}(x) \, = \,}
\\&&
\colour4colour{\cf^2}\*\Bigg\{
{{8\*(x+4)\*(x-1)}\over x}\*\H(1)-{{8\*(28\*x^2-36+3\*x+4\*x^3)}\over {15\*x}}\*\H(0)  -
\left({32\over 15}\*x^3-48\right)\*\Hh(0,0) - 16\*\H( 2)+
\nonumber\\&&
{{32\*(x+1)\*(x^4-x^3+x^2-6\*x+6)}\over {15\*x^2}}\*\Hh(-1,0) + \left(16+{32\over 15}\*x^3\right)\*\z2 -
{{8\*(x-1)\*(4\*x^2-27\*x-36)}\over {15\*x}}\Bigg\} +
\nonumber\\&&
\colour4colour{\cf\*\ca}\*\Bigg\{
-{{16\*(22-6\*x^2-21\*x+x^3 )}\over {3\*x}}\*\H(0)
+ {{16\*(x-1)\*(x^2-2\*x-29)}\over {3\*x}}\*\H(1) +
64\*\H(2)+
\nonumber\\&&
{{32\*(x+1)}\over x}\*\Hh(-1,0 )
-{{32\*(3\*x+4)}\over x}\*\Hh(0,0)
+ {{64\*(x-1)}\over x}\*\Hh(1,0)
-{{32\*(x-1)}\over x}\*\Hh(1,1) -
\nonumber\\&&
{{32\*(2\*x- 1)}\over x}\*\z2
+ {{32\*(x-1)\*(x^2-4\*x-14)}\over {3\*x}} \Bigg\}
\: \: ,\nonumber\\[2ex]
\lefteqn{ c_{A,\rm q}^{(2)}(x) - c_{T,\rm q}^{(2)}(x) \, = \,}
\\&&
\colour4colour{\cf^2}\*\Bigg\{
\pqq(-x)\*\Bigg((16\*\z2+16)\*\H(0)+48\*\Hhh(-1,0,0)-32\*\Hh(-1,2)+16\*\H(3) -
\nonumber\\&&
72\*\Hhh(0,0,0)+16\*\Hh(-2,0)+64\*\Hhh(-1,-1,0)-56\*\z3+64\*\H(-1)\*\z2\Bigg) -
{{2\*(153\*x+24-107\*x^2+24\*x^3)}\over {5\*x}}\*\H(0) +
\nonumber\\&&
((16-16\*x)\*\z2+18\*x-18)\*\H(1) +
(20+12\*x)\*\H(2)+(16-16\*x)\*\Hh(1,0)+(-8+8\*x)\*\Hh(1,1) +
\nonumber\\&&
(-16+16\*x)\*\Hh(-2,0) +
{{16\*(x+1)\*(3\*x^2+x+3)\*(x^2-3\*x+1)}\over {5\*x^2}}\*\Hh(-1,0) +
\nonumber\\&&
\left(-20-44\*x+16\*x^2-{48\over 5}\*x^3\right)\*\Hh(0,0) - (16-16\*x)\*\Hhh(1,0,0) +
\left(-16\*x^2-20\*x-28+{48\over 5}\*x^3\right)\*\z2 +
\nonumber\\&&
(16-16\*x)\*\z3 - {{(x-1)\*(48\*x^2-139\*x+48)}\over {5\*x}}\Bigg\} +
\nonumber\\&&
\colour4colour{\cf\*\ca}\*\Bigg\{
\pqq(-x)\*\Bigg((-8\*\z2-8)\*\H(0)+28\*\z3+36\*\Hhh(0,0,0)-8\*\H(3)-32\*\Hhh(-1,-1,0)+
16\*\Hh(-1,2)-
\nonumber\\&&
24\*\Hhh(-1,0,0) - 8\*\Hh(-2,0)-32\*\H(-1)\*\z2\Bigg) + {{4\*(18-104\*x^2+18\*x^3+161\*x)}\over {15\*x}}\*\H(0) +
\Bigg((-8+8\*x)\*\z2 -
\nonumber\\&&
{2\over 3}\*x+{2\over 3}\Bigg)\*\H(1) + (-8\*x-8)\*\H(2) + (8-8\*x)\*\Hh(-2,0) -
{{8\*(x+1)\*(3\*x^2+x+3)\*(x^2-3\*x+1)}\over {5\*x^2}}\*\Hh(-1,0) +
\nonumber\\&&
\left(16\*x+16-8\*x^2+{24\over 5}\*x^3\right)\*\Hh(0,0)+(8-8\*x)\*\Hhh(1,0,0) +
\left(8\*x^2+12\*x+12-{24\over 5}\*x^3\right)\*\z2 +
\nonumber\\&&
(-8+8\*x)\*\z3 + {{(x-1)\*(216\*x^2+157\*x+216)}\over {45\*x}} \Bigg\} +
\nonumber\\&&
\colour4colour{\cf\*\nf}\*\Bigg\{
\left(-{4\over 3}+{4\over 3}\*x\right)\*\H(0) + \left({4\over 3}-{4\over 3}\*x\right)\*\H(1) -
{38\over 9}\*x+{38\over 9} \Bigg\}
\: \: .\nonumber
\end{eqnarray}

\end{document}